\documentclass[reprint,amsmath,amssymb,aps,floatfix,pra]{revtex4-2}
\usepackage{amsmath}
\usepackage{hyperref}
\usepackage{graphicx}
\usepackage{bm}
\usepackage[normalem]{ulem}

\usepackage{amsbsy}
\usepackage{dcolumn}
\usepackage{color}
\usepackage{amssymb}
\usepackage{lipsum}
\usepackage{verbatim}
\usepackage{epsfig}
\usepackage[caption=false]{subfig}
\usepackage{feynmp-auto}
\newcommand{\be}{\begin{equation}}
\newcommand{\ee}{\end{equation}}
\newcommand{\bea}{\begin{eqnarray}}
\newcommand{\eea}{\end{eqnarray}}

\DeclareUnicodeCharacter{03B2}{\ensuremath{\beta}}
\DeclareUnicodeCharacter{03BD}{\ensuremath{\nu}}

\begin{document}
\title{Nonlocal nucleon-nucleus optical potentials from chiral effective field theory}
\author{L.\ M.\ Stahulak$^1$, J.\ W.\ Holt$^1$}
\affiliation{$^1$Cyclotron Institute and Dept.\ of Physics and Astronomy, Texas A\&M University, College Station, TX}  

\begin{abstract}
We investigate the nonlocality in microscopic optical potentials derived from chiral effective field theory. For this purpose we employ the Perey-Buck ansatz, which connects the energy dependence of purely local optical potentials to a Gaussian spatial nonlocality. We find that the dominant source of energy dependence in the microscopic real optical potential indeed arises from spatial nonlocalities, while the energy dependence associated with the microscopic imaginary optical potential is primarily a genuine time nonlocality. We present results for nonlocal nucleon-nucleus optical potentials for the calcium isotopic chain and study the dependence of the Woods-Saxon shape parameters on the isotopic number.
\end{abstract}

\maketitle

\section{Introduction}
\label{introduction}
Reaction studies involving exotic isotopes are important for an improved understanding of the nuclear force and the theoretical modeling of strongly interacting matter at the limits of stability \cite{Roussel-Chomaz2011}. Nuclear reactions are therefore central to many of the experimental programs at rare-isotope beam facilities, such as FRIB, which will explore structure properties and capture rates needed for nuclear astrophysics simulations and stockpile stewardship applications. One class of experiments of particular interest are neutron capture cross sections for heavy and neutron-rich nuclei, as these cross sections are vital for understanding r-process nucleosynthesis in exotic stellar sites \cite{Balantekin2014,Kasen2017}. Breakup reactions of exotic nuclei are also important for a variety of applications, and can be studied in rare-isotope beam facilities through both direct and surrogate experiments \cite{Bonaccorso2018}.
Knockout and transfer reactions involving highly isospin-asymmetric nuclei can probe the underlying nuclear structure but may require a modification of reaction theories that could be elucidated by experiments measuring their cross sections \cite{Hebborn062022}.

To connect the wealth of upcoming data to nuclear structure models and applications, it is essential to develop improved theoretical descriptions of nuclear scattering and reactions for medium-mass and heavy nuclei, especially those far from stability that may lie beyond the scope of phenomenological methods \cite{Goriely2007}. 
For most reaction studies involving medium-mass and heavy isotopes, the nuclear optical model potential (OMP) plays a key role by replacing the complicated many-body Hamiltonian with a simplified one-body average potential seen by the projectile. The most widely used and successful OMPs for target nuclei near the valley of stability are generally constructed phenomenologically \cite{Hebborn102022}. While these phenomenological OMPs for elastic and inelastic scattering are quite common, they do not directly reflect the underlying microscopic physics at work. In addition, phenomenological OMPs require a significant amount of scattering data and therefore may not extrapolate well beyond the valley of stability where data is scarce \cite{Koning2003}. Although reevaluations of experimental uncertainties and outlier rejection can mitigate this, there are inherent problems in trying to apply phenomenological potentials across the nuclear chart \cite{Pruitt2022}. 

Analysis has shown that the uncertainty introduced from phenomenological optical potentials varies considerably depending on the characteristics of the nucleus.
Since nuclei near the dripline often show novel behaviors, such as changing magic numbers and halo features, these systems may pose challenges for phenomenological potentials extrapolated from stable regions \cite{Jonson2004}. In particular, halo nuclei ground states contain nucleons separated from the core beyond the range of the nuclear interaction. Knockout reactions of these systems are expected to produce interesting information on the structure of these exotic nuclei \cite{Hebborn2020,Hebborn092021}. These reactions are important in the production of many exotic nuclei of interest, and therefore the theoretical methods for calculating the relevant cross sections for these processes must be carefully scrutinized \cite{Hebborn062022}.

In contrast to phenomenological OMPs, microscopic OMPs are not fitted directly to experimental scattering data. If the underlying theory uncertainties are under control, microscopic approaches may therefore be more suitable to extrapolating into regions where data is scarce. Modern approaches to constructing microscopic optical potentials include ab initio methods using particle and hole Green's functions \cite{Rotureau2017,Rotureau2018,Idini2017} and calculations of the nucleon self-energy in nuclear matter, which is linked directly to the OMP \cite{Dickhoff2019}. Ab initio approaches are generally limited to lighter nuclei and low projectile energies. The infinite nuclear matter approach \cite{Jeukenne1977}, on the other hand, can be applied across a large range of isotopes and projectile energies, but it is subject to uncontrolled systematic uncertainties associated with the local density approximation. All microscopic approaches are well suited to capture the isovector components of the OMP, which are vital to understanding scattering involving highly asymmetric nuclei \cite{Weppner2009}. The isovector dependence of the imaginary part of the optical potential also has been suggested to have significant implications for r-process nucleosynthesis \cite{Goriely2007}. Theoretical tools to calculate neutron capture on exotic nuclei will therefore be very important to direct future nuclear astrophysics research. 

In the present work, we focus on the infinite nuclear matter approach, where the nucleon self-energy is first computed in homogeneous nuclear matter at varying density and isospin asymmetry. One then combines the infinite matter self-energy with a target nuclear density distribution and the local density approximation to construct a nucleon-nucleus optical potential. Recently, Whitehead, Lim, and Holt (WLH) \cite{Whitehead2019,Whitehead2020,Whitehead2021} started from microscopic two-body and three-body nuclear forces derived with the framework of chiral effective field (ChEFT) \cite{epelbaum09rmp,machleidt11} to construct a microscopic global nucleon-nucleus optical potential suitable for a wide range of target nuclei and projectile energies up to about $E\simeq 150$\,MeV. ChEFT enables the estimation of many-body uncertainties that arise from the choice of nuclear interaction through (i) the EFT truncation order \cite{Young2003,Furnstahl2015,Melendez2019,Wesolowski2022}, (ii) the fitting of the short-distance contact interactions in the theory \cite{Svensson2022,VanGoffrier2025}, and (iii) the resolution scale encoded in the momentum-space cutoff $\Lambda$ \cite{Machleidt2011,Tews2018,Tews2025}. Within theoretical uncertainties, the WLH predictions for single-nucleon elastic differential scattering cross sections and analyzing powers for stable nuclei were shown to match existing data quite well. In the present work, we employ the same two-body and three-body chiral nuclear forces that were used in Ref.\ \cite{Whitehead2021} to construct the WLH global potential.

Normally, global optical potentials based on the nuclear matter approach are calculated ``on-shell'', meaning that the separate energy and momentum dependence of the self-energy $\Sigma(q,E)$ is replaced with a single energy dependence resulting from the on-shell condition
\begin{equation}
E(q) = \frac{q^2}{2M} + \mathrm{Re}\Sigma(q,E(q))
    \label{Eq:SelfConsistentE},
\end{equation}
where $M$ is the free-space nucleon mass. This constraint leads to a local but energy-dependent optical potential, which is relatively simple to implement in a wide variety of open-source reaction codes. The practice of combining the separate momentum and energy dependencies is technically valid only for elastic scattering, where no energy is transferred \cite{Titus2014,Titus2016,Ross2015}. However, optical model potentials are routinely used to study inelastic reaction processes that may be especially sensitive to specific underlying structural features, such as the location of certain excited states, and which therefore may be more sensitive to differences between nonlocal and energy-dependent local optical potentials.
For instance, predicted projectile wavefunctions differ significantly in the surface and exterior regions of the nucleus when comparing calculations using local vs. nonlocal optical potentials. These differences have been found to have important consequences in transfer reactions \cite{Titus2014,Titus2016}. In particular, strong nonlocalities in deuteron scattering states suppress the wavefunction in the interior of the nucleus, making
differences in the periphery more important \cite{Titus2016}.

Given that the range of applications where useful optical potentials are desired has expanded beyond elastic scattering \cite{Arellano2019,Deltuva2009,Odell2024,Timofeyuk2013,Perrotta2025}, it is necessary to understand the effects of making the on-shell approximation, effectively absorbing all spatial and temporal nonlocalities into a single energy-dependent local optical potential \cite{Arellano2018,Arellano2022}. Much work \cite{Ripka1963,Giannini1976} has been done to characterize the nonlocality of the optical potential obtained from microscopic nucleon-nucleon interactions. The primary contributions to the spatial nonlocality come from exchange effects at first order in perturbation theory, while temporal nonlocalities arise at second order and beyond due to virtual transitions to excited states with nonzero decay widths.

For elastic scattering, one way to construct a nonlocal optical potential from an equivalent energy-dependent but local optical potential is via the Perey-Buck approach \cite{Perey1962}, where the optical potential is assumed to factorize into separate local and nonlocal terms and where the nonlocal part is further assumed to have a Gaussian shape. In the present work, we start from local, energy-dependent microscopic optical potentials constructed from chiral nuclear forces to derive associated nonlocal, energy-independent optical potentials using the Perey-Buck approach. We find that the Perey-Buck ansatz adequately captures the nonlocal aspects of microscopic nucleon-nucleus optical potentials up to projectile energies of $E\sim 100$\,MeV, beyond which the Perey-Buck assumption appears to break down.

The paper is organized as follows. In Section \ref{Sec:Methods} we demonstrate how to obtain nonlocal optical potentials by combining the nuclear matter approach with the Perey-Buck ansatz. In Section \ref{Sec:Results} we present and discuss the results of our analysis for OMPs derived from chiral effective field theory, including the parameters of the nonlocal OMPs for calcium isotopes. In Section \ref{Sec:Conclusion} we summarize the work and discuss directions for further research.
\section{Nonlocal microscopic optical potentials from the Perey-Buck Ansatz}
\label{Sec:Methods}
\begin{figure}
    \centering
    \includegraphics[width=0.3\linewidth,page=1]{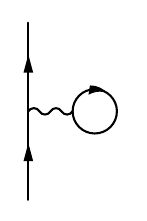}
    \includegraphics[width=0.3\linewidth,page=2]{plots/MBPTDiags.pdf}
    \includegraphics[width=0.3\linewidth,page=3]{plots/MBPTDiags.pdf}
    \caption{Perturbation theory diagrams representing the first- and second-order contributions to the self-energy. Wavy lines represent the ChEFT nucleon-nucleon interaction.}
    \label{Fig:MBPTDiagrams}
\end{figure}

\subsection{Nuclear matter optical potentials from chiral effective field theory}
\label{SubSec:ChiEFTSelfEnergy}

One approach to calculate nucleon-nucleus optical potentials is to start from the self-energy of a nucleon propagating in homogeneous nuclear matter with fixed density and proton fraction. By varying the density and composition of nuclear matter over the range of conditions encountered in finite nuclei, one can then use the local density approximation (LDA) \cite{Jeukenne1977,Brieva1977} or the improved local density approximation (ILDA) \cite{Bauge1998} to relate the nucleon-nucleus optical potential at a given separation distance to the local values of density and isospin asymmetry \cite{Bauge1998}. This approach was used to derive the WLH microscopic global optical potential \cite{Whitehead2021} and will also be the basis for the approach in our study.

The self-energy in nuclear matter is obtained by performing a many-body perturbation theory (MBPT) expansion in powers of the nuclear potential. The first- and second-order contributions to the self-energy from MBPT are shown diagrammatically in Figure \ref{Fig:MBPTDiagrams}. Numerically, these are evaluated using the following sums over states in nuclear matter:
\begin{align}
    \Sigma^{(1)}&(q,E;k_f) = \sum_1\langle \vec{q} \vec{h}_1 ss_1tt_1|\bar{V}^{\mathrm{eff}}_{2N}|\vec{q}\vec{h}_1ss_1tt_1\rangle n_1,
    \label{Eq:Sigma1}
\end{align}
\begin{align}
    \nonumber \Sigma^{(2a)}&(q,E;k_f) = \\
    &\frac{1}{2}\sum_{123}\frac{|\langle\vec{p_1}\vec{p_3}s_1s_3t_1t_3|\bar{V}^{\mathrm{eff}}_{2N}|\vec{q}\vec{h}_2ss_2tt_2\rangle|^2}{E + E_2 - E_1 - E_3 + i\eta} \bar{n}_1 n_2 \bar{n}_3,
    \label{Eq:Sigma2a}
\end{align}
\begin{align}
    \nonumber \Sigma^{(2b)}&(q,E;k_f) = \\
    &\frac{1}{2}\sum_{123}\frac{|\langle\vec{h}_1\vec{h}_3s_1s_3t_1t_3|\bar{V}^{\mathrm{eff}}_{2N}|\vec{q}\vec{p}_2ss_2tt_2\rangle|^2}{E + E_2 - E_1 - E_3 - i\eta} n_1 \bar{n}_2 n_3.
    \label{Eq:Sigma2b}
\end{align}
Since these expressions depend explicitly on the momentum $q$ and energy $E$, Fourier transforms lead to associated spatial and temporal nonlocalities, respectively, in the nucleon self-energy and hence the optical potential. The above equations are summed over the momenta $\vec p$, spins $s$, and isospins $t$ of the intermediate particle and hole states. The function $n_i = \theta(k_f-|\vec{h}_i|)$ is the zero temperature occupation probability, and $\bar{n}_i = 1-n_i$ selects particle states above the Fermi momentum. The operator $\bar{V}^{\mathrm{eff}}_{2N}$ is the antisymmetrized effective two-nucleon interaction from ChEFT. The single-particle energies $E_i$ in Eqs.\ \eqref{Eq:Sigma2a} and \eqref{Eq:Sigma2b} are solved self-consistently in the process of calculating the self-energy, using Eq.\ \eqref{Eq:SelfConsistentE}. This dispersion relation effectively transforms the separate energy and momentum dependencies of the self-energy into a single combined energy dependence. However, this relation relies on the conservation of momentum for the ingoing and outgoing nucleon in the self-energy, but because momentum is transferred in inelastic processes such as knockout or breakup, Eq.\ \eqref{Eq:SelfConsistentE} would not hold in these cases. The nature and degree of the discrepancies caused by the breakdown of these assumptions should be investigated, especially in the case of exotic nuclei. We are therefore interested in characterizing the effects these assumptions have on any calculated observables.

The nucleon self-energy in nuclear matter can be separated into its real and imaginary parts:
\begin{equation}
    V_i(E;k_f^p,k_f^n) = U_i(E;k_f^p,k_f^n) + iW_i(E;k_f^p,k_f^n),
\end{equation}
where the index $i = \{p,n\}$ indicates either a propagating proton or neutron respectively, and 
\begin{align}
    U_i(E;k_f^p,k_f^n)& = \mathrm{Re}\Sigma_i(q,E(q);k_f^p,k_f^n),\\
    W_i(E;k_f^p,k_f^n)& = \frac{M_i^{k*}}{M} \mathrm{Im}\Sigma_i(q,E(q);k_f^p,k_f^n),
\end{align}
where the imaginary part of the optical potential is scaled by the effective $k$-mass
\begin{equation}
    \frac{M_i^{k*}}{M} = \bigg(1+\frac{M}{k}\frac{\partial}{\partial k} U_i(k,E(k)) \bigg)^{-1}
\end{equation}
in order to account for the imaginary part missing in Eq. \eqref{Eq:SelfConsistentE} \cite{JWNegele1981,SFANTONI1981}.

Because the spin-orbit interaction vanishes in infinite nuclear matter, the above procedure accounts for only the real and imaginary volume terms and the imaginary surface term of the optical potential. For the spin-orbit interaction, we employ the density matrix expansion of Negele and Vautherin \cite{Bogner2009,Negele1972}. Starting from the density matrix 
\begin{equation}
\rho(\vec{r}_1\sigma_1\tau_1;\vec{r}_2\sigma_2\tau_2)=\sum_\alpha\Psi^*_\alpha(\vec{r}_2\sigma_2\tau_2)\Psi_\alpha(\vec{r}_1\sigma_1\tau_1),
\label{dm}
\end{equation}
where $\Psi_\alpha$ are the single-particle orbitals in the nucleus, an associated nuclear energy-density functional at the Hartree-Fock level
\begin{align}
    \nonumber \mathcal{E}[\rho,\tau,\vec{J}\,]&=\rho\bar{E}(\rho)+\bigg[\tau-\frac{3}{5}\rho k_f^2\bigg]\bigg[\frac{1}{2M_N}+F_\tau(\rho)\bigg]\\
    & \hspace{-.1in}+(\vec{\nabla}\rho)^2F_\nabla(\rho)+\vec{\nabla}\rho\cdot\vec{J}F_{SO}(\rho)+\vec{J}^{\,2}F_J(\rho)
    \label{edf}
\end{align}
can be extracted by first separating Eq.\ \eqref{dm} into relative and center-of-mass coordinates and then expanding up to second order in spatial gradients. In Eq.\ \eqref{edf}, $\tau$ represents the local kinetic energy density, $\rho$ is the local nucleon density, and $\vec J$ is the local spin-orbit density, while $\{ \bar E, F_\tau, F_\nabla, F_{SO}, F_J \}$ are the scalar strength functions that can be extracted from the underlying microscopic nuclear interaction. Of particular interest is the spin-orbit strength $F_{SO}$, which is extracted and employed as the energy-independent optical potential corresponding to the spin-orbit interaction.

In this work, we choose for the nuclear interaction in Eqs.\ \eqref{Eq:Sigma1}-\eqref{Eq:Sigma2b} microscopic nuclear two-body and three-body chiral nuclear interactions. All calculations were performed using nucleon-nucleon interactions up to N3LO (next-to-next-to-next-to-leading order) and three-body forces at N2LO, all with a momentum-space cutoff of $\Lambda$ = 450 MeV.

\subsection{Perey-Buck Nonlocal Optical Potentials}
\label{SubSec:Perey-BuckEquiv}
In principle, it is possible to study the independent momentum and energy dependence of the nucleon optical potential in nuclear matter through the off-shell self-energy. An explicit Fourier transform to coordinate space would then yield a nonlocal, energy-dependent optical potential in nuclear matter. However, for practical implementation into nucleon-nucleus reaction codes, it is useful to have a parameterized form for the spatial nonlocality. We therefore propose to investigate the energy dependence of the microscopic optical potential in nuclear matter via the Perey-Buck ansatz and then employ the improved local density approximation to construct parameterized nonlocal nucleon-nucleus optical potentials.

Perey and Buck \cite{Perey1962} started by writing the nonlocal nucleon-nucleus optical potential in the factorized form
\begin{equation}
    U(\vec{r},\vec{r}\,') = U_N(\tfrac{1}{2}|\vec{r}+\vec{r}\,'|) H(|\vec{r}-\vec{r}\,'|),
    \label{PBKernel}
\end{equation}
where both $U_N$ and $H$ are energy independent.
The term $U_N(\tfrac{1}{2}|\vec{r}+\vec{r}\,'|)$, which would be absent in the case of infinite homogeneous nuclear matter, is assumed to have the usual Woods-Saxon form, whereas the nonlocal term $H(|\vec{r}-\vec{r}\,'|)$ is assumed to have a Gaussian form given by
\begin{equation}
    H(|\vec{r}-\vec{r}\,'|) = \frac{1}{\pi^{3/2}\beta^3}\mathrm{exp}\bigg[-\bigg(\frac{|\vec{r}-\vec{r}\,'|}{\beta}\bigg)^2\bigg],
    \label{PBNonlocality}
\end{equation}
where $\beta$ is the nonlocality length scale parameter. Perey and Buck then derived a correspondence between energy-dependent local optical potentials and energy-independent nonlocal optical potentials by noting that the nucleon-nucleus elastic scattering wavefunctions for both cases differ very little outside of the nucleus itself, and making the assumption that well within the nucleus the local potential is essentially constant \cite{Perey1962}. One of the principal results from the Perey-Buck analysis is the following relation connecting the local and nonlocal optical potential depths:

\begin{equation}
    U_L(E) \cdot \mathrm{exp}\Bigg[\frac{M\beta^2}{2}(E-U_L(E))\Bigg] = U_N,
    \label{Eq:Perey-BuckEquiv}
\end{equation}
where $U_L(E)$ is the depth of the equivalent energy-dependent
local optical potential. We test this assumption using the results of $U_L(E)$ obtained from the on-shell self-energy up to second order in perturbation theory from chiral nuclear forces. For nuclear matter at a given density and isospin asymmetry, we determine the energy-independent values of $U_N$ and $\beta$ that minimize the least squares error of Eq.\ \eqref{Eq:Perey-BuckEquiv} taken over a fixed range $0<E<E_{\rm max}$ of scattering energies. We also investigate different value of $E_{\rm max}$ to determine at what energies the relation in Eq.\ \eqref{Eq:Perey-BuckEquiv} breaks down.

\subsection{Improved Local Density Approximation}
\label{SubSec:ILDA}
After $\beta$ and $U_N$ are determined over a range of densities and isospin asymmetries, we fold $U_N$ and $\beta$ with the nucleus density distribution in order to obtain the equivalent values at each point in the nucleus according to the improved local density approximation (ILDA):
\begin{align}
    \nonumber U_{\mathrm{ILDA}}(r) &= \frac{1}{(t\sqrt{\pi})^3}\int U_{\mathrm{LDA}}(r')e^{\frac{-|\vec{r}-\vec{r}'|^2}{t^2}}d^3r'\\
    \beta_{\mathrm{ILDA}}(r) &= \frac{1}{(t\sqrt{\pi})^3}\int \beta_{\mathrm{LDA}}(r')e^{\frac{-|\vec{r}-\vec{r}'|^2}{t^2}}d^3r',
\end{align}
where the optical potential parameters at a given distance from the nucleus depend only on the local conditions within a finite range of that point, with the range parameter $t$ chosen to be characteristic of the range of the nuclear interaction; in this work, the value $t = 1.2 \mathrm{ fm}$ was used. Thus, the radial dependence for $\beta$ and $U_N$ is obtained by interpolating the discrete grid of density and isospin-asymmetry points onto a radial map of the local density and isospin asymmetry values for a particular nucleus.

The nucleus density distributions used in this work are derived from a Skyrme effective interaction \cite{Lim2017,Whitehead2019} that was fitted to reproduce the empirical binding energies of finite nuclei as well as the nuclear matter equation of state from the N3LO-450 chiral potential that was used to generate the nucleon self-energies. The employed Skyrme mean field model has additional density-dependent interaction terms to better match the low-density equation of state from ChEFT.

\section{Results}
\label{Sec:Results}

\subsection{Testing the Perey-Buck Equivalence}
\label{SubSec:PBNonlocalResults}

\begin{figure*}[t]
    \subfloat{
     \centering
     \includegraphics[width=0.46\textwidth]{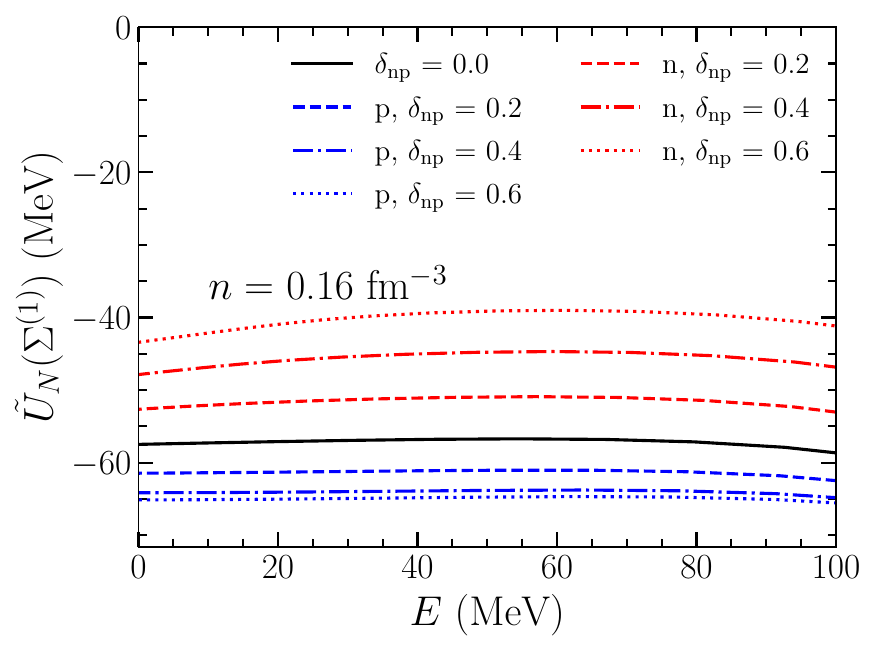}
     \label{Fig:1stAnd2ndOrderEDepUN}
   }\hfill
   \subfloat{
     \centering
     \includegraphics[width=0.46\textwidth]{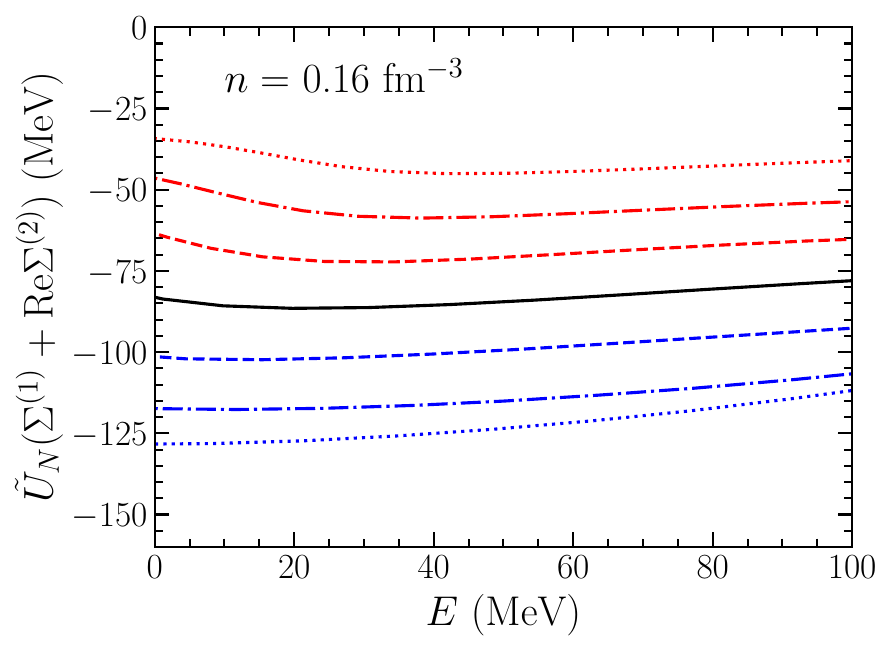}
     \label{Fig:1stOrderEDepUN}
   }
   \caption{Energy dependence of the Perey-Buck nonlocality function $\tilde U_N$ in Eq.\ \eqref{Eq:Perey-BuckEquiv} for protons and neutrons in nuclear matter at density $n$ = 0.16 fm$^{-1}$ and varying isospin asymmetry. The left panel is calculated from first-order corrections to the self-energy, while the right panel includes also second-order perturbative corrections to the self-energy.}
   \label{Fig:EDepUN}
\end{figure*}

Once the on-shell, energy-dependent self-energy is calculated up to second order in perturbation theory for a wide range of densities and isospin asymmetries, we parameterize the energy dependence via the Perey-Buck nonlocality equation (Eq.\ \eqref{Eq:Perey-BuckEquiv}). Values for the parameters $\beta$ and $U_N$ were determined such that the real part of the self-energy exhibited a minimized variation in the potential depth $U_N$ with the total projectile energy; this was done over the range 0-100 MeV for the projectile energy, minimizing the function
\begin{align}
    \nonumber J(U_N,\beta) = &\\ \sum_i \bigg( U_L(E_i) \cdot &\mathrm{exp}\Bigg[\frac{M\beta^2}{2}(E_i-U_L(E_i))\Bigg] - U_N\bigg)^2,
    \label{Eq:PBFitCostFunc}
\end{align}
where $i$ indexes discrete points within the aforementioned energy range. This was repeated for different values of the self-energy calculated across a range of densities and isospin asymmetries, constructing a discrete grid of pairs of the $U_N$ and $\beta$ parameters covering the typical conditions inside a nucleus.

At leading order in perturbation theory, there is no explicit energy dependence in the optical potential. In other words, the entirety of the obtained energy dependence just comes from the explicit momentum dependence combined with the on-shell approximation, Eq. \eqref{Eq:SelfConsistentE}. In the left panel of Figure \ref{Fig:EDepUN}, we show the energy dependence of the fitted nonlocal potential strength for protons (blue) and neutrons (red) from the real part of the 1st-order contribution to the self-energy ($U_N(\Sigma^{(1)})$). Specifically, we plot \begin{align*}
    \tilde U_N(E) = U_L(E) \cdot \mathrm{exp} \Bigg[\frac{M\beta^2}{2}(E-U_L(E))\Bigg].
\end{align*}
The density is fixed at $n=n_0 = 0.16$\,fm$^{-3}$ and the isospin asymmetry is varied over the values $\delta_{np}=0,0.2,0.4,0.6$. We see from the left panel of Figure \ref{Fig:EDepUN} that the quantity $\tilde U_N(E)$ can be made to vary minimally over the energy range $0 < E < 100$\,MeV. This holds for both protons and neutrons as well as for all values of the isospin asymmetry that we considered. This demonstrates that the Perey-Buck ansatz is well justified for optical potentials derived from ChEFT. In addition, we see that at low isospin asymmetries, the dependence of $U_N$ on $\delta_{np}$ follows the so-called Lane form: 
\begin{equation}
U_N(\delta_{np}) = U_N(0) + U_N^I \delta_{np} \tau_z,
\label{lane}
\end{equation}
where $U_N^I$ is the isovector part of the optical potential and $\tau_z$ is the isospin projection operator for the projectile. At higher values of the isospin asymmetry, $\delta_{np} \sim 0.6$, however, it appears that the Lane form breaks down, as observed in previous work \cite{Holt2016}. 

\begin{figure*}
    \subfloat{
        \centering
        \includegraphics[width=0.46\linewidth]{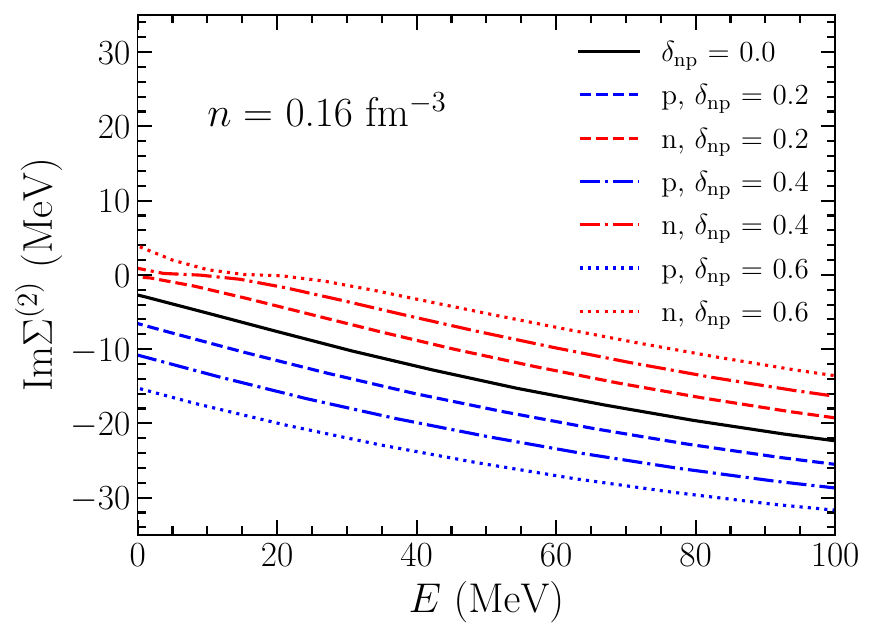}
    }\hfill
    \subfloat{
        \centering
        \includegraphics[width=0.46\linewidth]{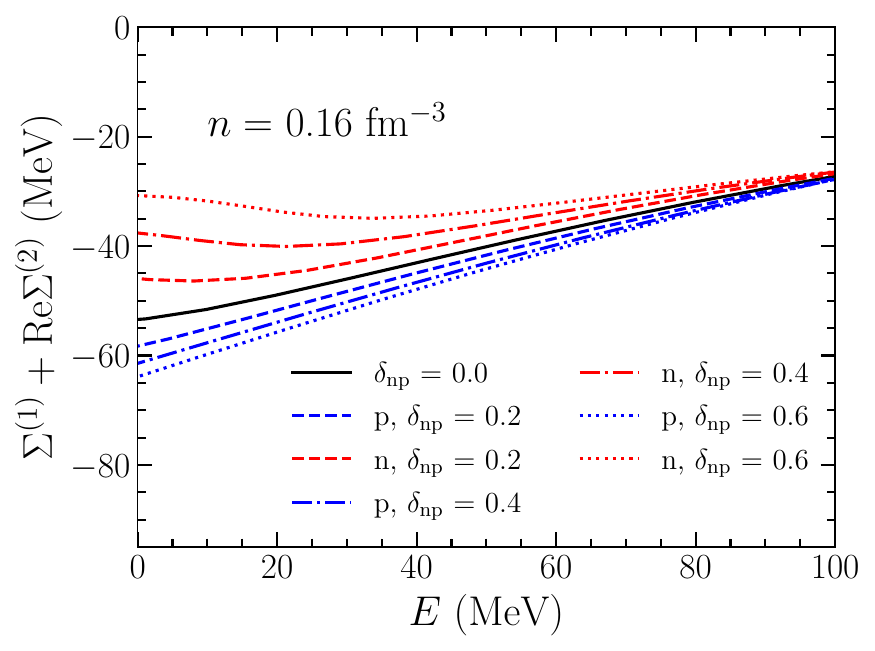}
    }
    \caption{ Energy dependence of Im$(\Sigma)$ and Re$(\Sigma)$ for $n = 0.16 \mathrm{ fm}^{-3}$. }
    \label{Fig:ImagEDep}
\end{figure*}

At second order in perturbation theory, the real part of the optical potential will have separate energy and momentum dependencies. As observed in previous works \cite{Holt2013,Holt2016}, however, the largest energy dependence occurs close to the Fermi surface, which lies at negative single-particle energies. It is therefore of interest to investigate to what extent an energy-independent but spatially nonlocal (i.e., momentum dependent) optical potential can characterize the microscopic results from chiral effective field theory in the strictly positive energy regime. In the right panel of Figure \ref{Fig:EDepUN}, we plot the quantity $\tilde U_N(E)$ for the combined input $\Sigma^{(1)} + \Sigma^{(2)}$. Again, we consider nuclear matter at the discrete values of the isospin asymmetry $\delta_{np}=0,0.2,0.4,0.6$ and consider both proton and neutron projectiles. We find that overall, the Perey-Buck ansatz can capture the results from microscopic many-body theory, though there is a somewhat larger variation in the energy compared to the left panel of Figure \ref{Fig:EDepUN} that includes $\Sigma^{(1)}$ alone. One also observes in the right panel that in most cases there is an increase in $\tilde U_N(\Sigma^{(1)}+\Sigma^{(2)})$ as $E \rightarrow0$, which stems from the significant energy dependence of $\Sigma^{(2)}$ around the (negative) Fermi energy.

From the large spacing of proton and neutron results in the right panel of Figure \ref{Fig:EDepUN} compared to the left panel, we see that the inclusion of the second-order self-energy leads to an isovector optical potential contribution that is at least twice as large across all energies compared to when $\Sigma^{(1)}$ alone is included. This is consistent with previous observations \cite{Holt:2011eg,Holt:2011yj} of the isospin-asymmetry dependence of the nuclear equation of state, where nuclear two-body forces at second-order in MBPT contribute significantly to the isospin asymmetry energy. Although the results in Figure \ref{Fig:EDepUN} are obtained for the specific density $n=n_0$, the general trends we have observed are also seen across all values for the density that were studied in this work, including number densities between 0.01 and 0.20 fm$^{-3}$. Finally, during our analysis we found that the applicability of the Perey-Buck energy dependence seems to fail beyond 100 MeV for both the first-order MBPT contribution alone and the combined first- and second-order results for the self-energy. It was not possible to obtain a satisfactory fit for $U_N$ and $\beta$ while including data beyond this range.

No part of the imaginary contribution $W_L(E)$ could be found to match the Perey-Buck energy dependence in Eq.\ \eqref{Eq:Perey-BuckEquiv}. In the left panel of Figure \ref{Fig:ImagEDep}, we show the energy dependence of the imaginary part of the second-order on-shell self-energy in nuclear matter at saturation density and for different isospin asymmetries. One sees generically that the magnitude of the imaginary part increases, but at a rate that is much less than the increase in the energy. This is the key point, because from inspection of Eq.\ \eqref{Eq:Perey-BuckEquiv}, we see that in order for the right-hand side to remain constant, the exponential term on the left-hand side needs to decrease in magnitude to counteract the increase in $W_L(E)$. But since $E$ grows more quickly than $W_L(\Sigma^{(2)})$, both factors on the left-hand side of Eq.\ \eqref{Eq:Perey-BuckEquiv} increase with the energy, behavior which is inconsistent with the Perey-Buck ansatz. On the other hand, as can be seen in the right panel of Figure \ref{Fig:ImagEDep} the magnitude of the real part of the optical potential $U_L$ decreases with the energy and at a rate $d|U|/dE <1$. Thus, the exponential term on the left-hand side of Eq.\ \eqref{Eq:Perey-BuckEquiv} naturally increases in magnitude and fitting to the Perey-Buck form works well.

\begin{figure*}[t]
    \subfloat{
        \centering
        \includegraphics[width=0.46\linewidth]{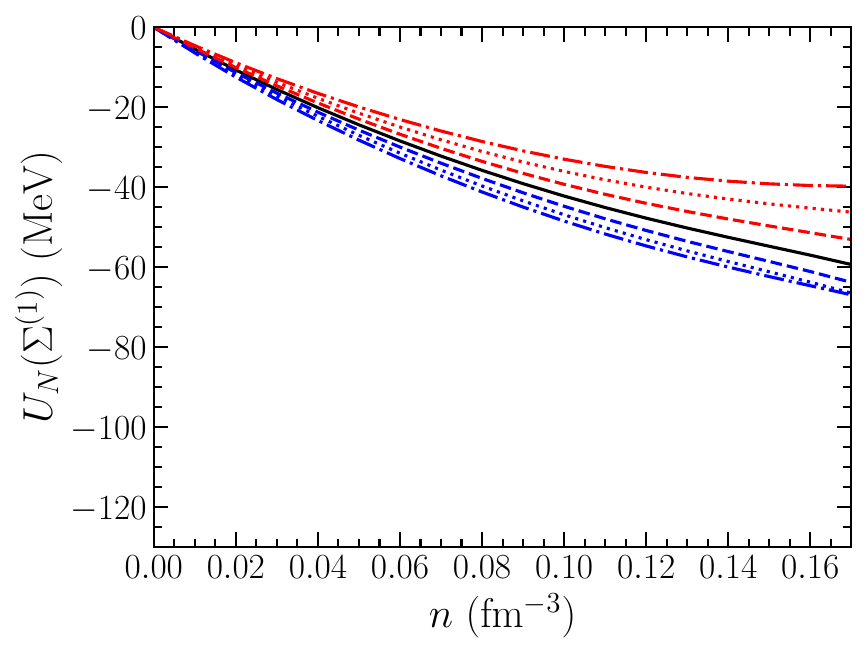}
        \label{Fig:1stOrderRhoDepUN}
    }\hfill
    \subfloat{
        \centering
        \includegraphics[width=0.46\linewidth]{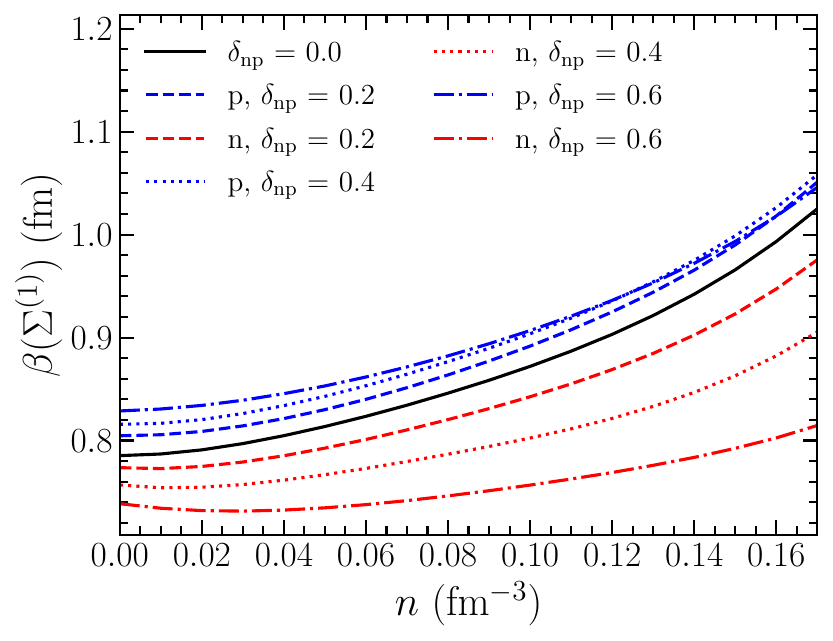}
        \label{Fig:1stOrderRhoDepBeta}
    }
    \caption{Density dependence of $U_N(\Sigma^{(1)})$ (left) and $\beta(\Sigma^{(1)})$ (right) for protons (blue) and neutrons (red) in nuclear matter at varying isospin asymmetry $\delta_{np}$.}
    \label{Fig:1stOrderRhoDep}
\end{figure*}
\begin{figure*}[t]
    \subfloat{
        \centering
        \includegraphics[width=0.46\textwidth]{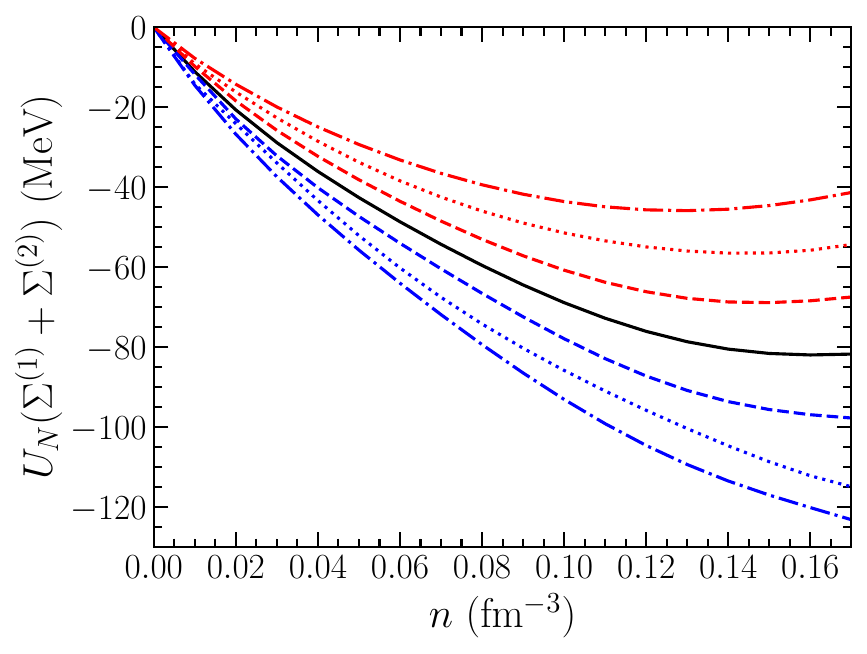}
    }\hfill
    \subfloat{
        \centering
        \includegraphics[width=0.46\textwidth]{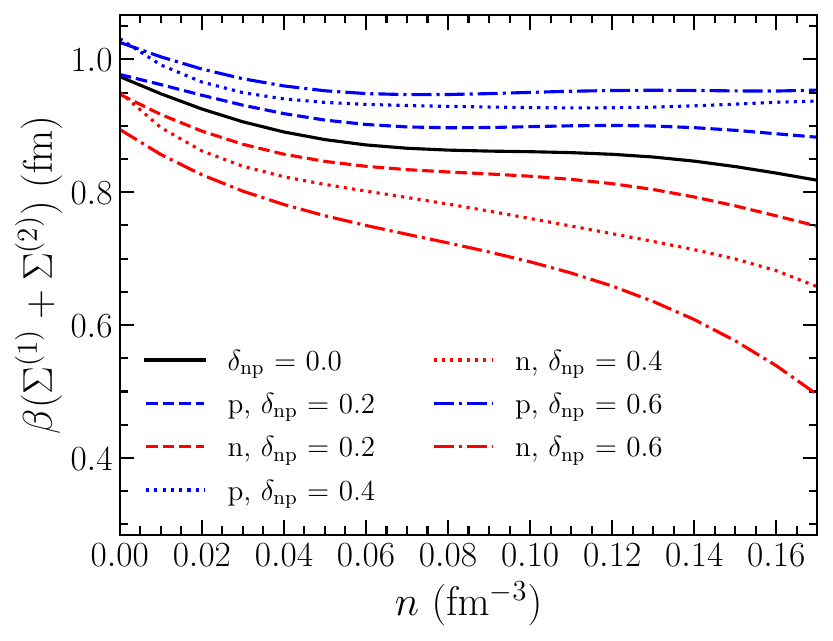}
    }
    \caption{Density dependence of $U_N(\Sigma^{(1)}+\Sigma^{(2)})$ (left) and $\beta(\Sigma^{(1)}+\Sigma^{(2)})$ (right) for protons (blue) and neutrons (red) in nuclear matter at varying isospin asymmetry $\delta_{np}$. }
    \label{Fig:1stAnd2ndOrderRhoDep}
\end{figure*}

\subsection{Density and isospin asymmetry dependence of the nonlocal optical potential parameters}
\label{SubSec:DensityDep}

\begin{figure*}[t]
    \subfloat{
    \centering
    \includegraphics[width=0.46\linewidth]{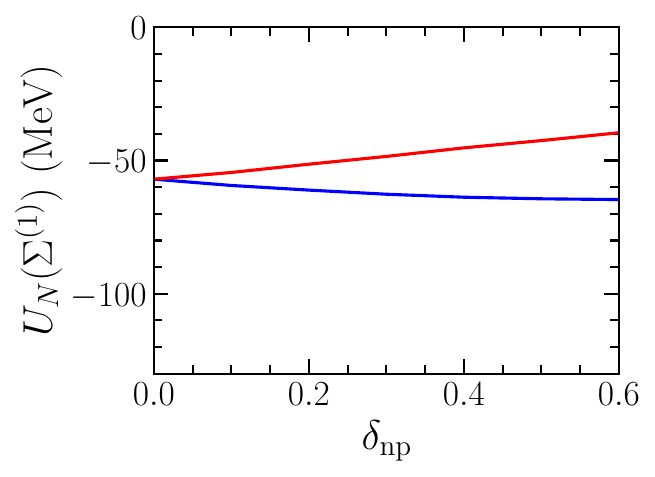}
    }
    \subfloat{
    \centering
        \includegraphics[width=0.46\linewidth]{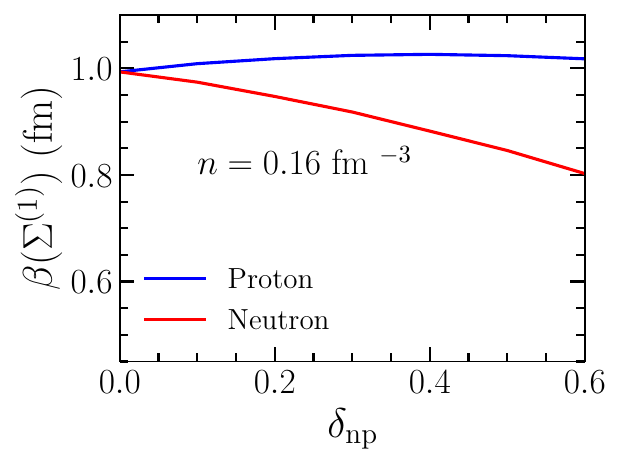}
    }
    
    \centering
    \subfloat{
    \centering
        \includegraphics[width=0.46\linewidth]{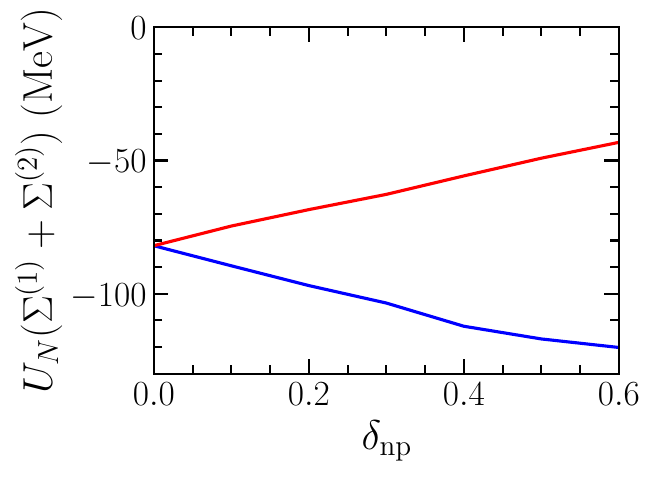}
    }
    \subfloat{
    \centering
        \includegraphics[width=0.46\linewidth]{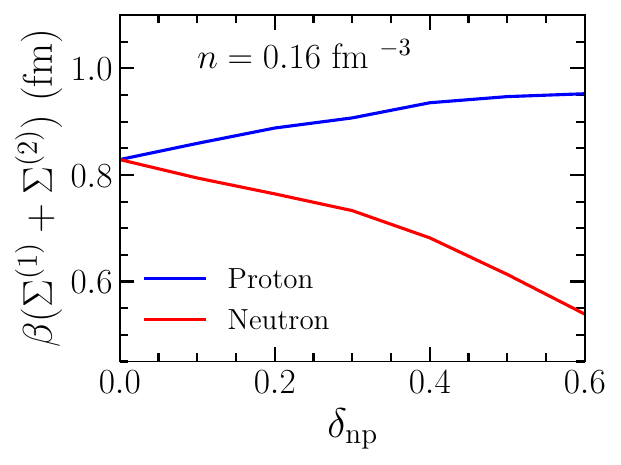}
    }

    \caption{Isospin asymmetry dependence of the proton (blue) and neutron (red) Perey-Buck nonlocality parameters  $U_N(\Sigma^{(1)})$, $\beta(\Sigma^{(1)})$, $U_N(\Sigma^{(1)}+\Sigma^{(2)})$, and $\beta(\Sigma^{(1)}+\Sigma^{(2)})$ in nuclear matter at density $n=n_0$.}
    \label{Fig:AsymmetryDep}
\end{figure*}

In the left and right panels of Figure \ref{Fig:1stOrderRhoDep}, we show the density dependence of the best fitted values of $U_N(\Sigma^{(1)})$ and $\beta(\Sigma^{(1)})$ for protons and neutrons in homogeneous nuclear matter as a function of the density and selected isospin asymmetries of $\delta_{np}=0, 0.2, 0.4, 0.6$. As expected, the nonlocal potential depth $U_N$ increases with increasing density, and protons experience a stronger optical potential than neutrons in neutron-rich matter. In addition, we find that the nonlocality parameter $\beta$ also increases with the density. This is associated with the fact that the effective nucleon mass strongly decreases with increasing density \cite{Holt2011} at first-order in perturbation theory. Microscopic nuclear forces also generically predict a smaller effective mass for protons in neutron-rich matter compared to neutrons. This is reflected in the larger proton nonlocality parameter $\beta$ compared to neutrons, as seen in the right panel of Figure \ref{Fig:1stOrderRhoDep}.

In Figure \ref{Fig:1stAnd2ndOrderRhoDep} we show the dependence of $U_N(\Sigma^{(1)}+\Sigma^{(2)})$ and $\beta(\Sigma^{(1)}+\Sigma^{(2)})$ on the nucleon number density and isospin asymmetry. Qualitatively, we find that the nonlocal potential depth $U_N$ exhibits similar properties to those of $U_N(\Sigma^{(1)})$ shown in the left panel of Figure \ref{Fig:1stOrderRhoDep}, except that the depth grows stronger with the density and the associated isovector optical potential is larger. However, comparing $\beta(\Sigma^{(1)}+\Sigma^{(2)})$ to the results in the right panel of Figure \ref{Fig:1stOrderRhoDep}, we see significant differences. In particular, the nonlocality length parameter $\beta$ shows a much weaker dependence on the density, and the neutron nonlocality parameter even decreases with density for most values of the isospin asymmetry. Indeed, previous calculations \cite{Holt2018} of the nucleon effective mass within Landau's Fermi liquid theory show that the inclusion of second-order perturbation theory diagrams mitigates the rapid decline in the effective mass due to first-order effects. One notices, however, that the proton nonlocality range parameter $\beta$ remains always larger than the neutron nonlocality parameter across all densities. One can infer that the proton effective mass remains below the neutron effective mass in neutron-rich matter.

\begin{figure*}[t]
    \subfloat{
     \centering
     \includegraphics[width=0.46\linewidth]{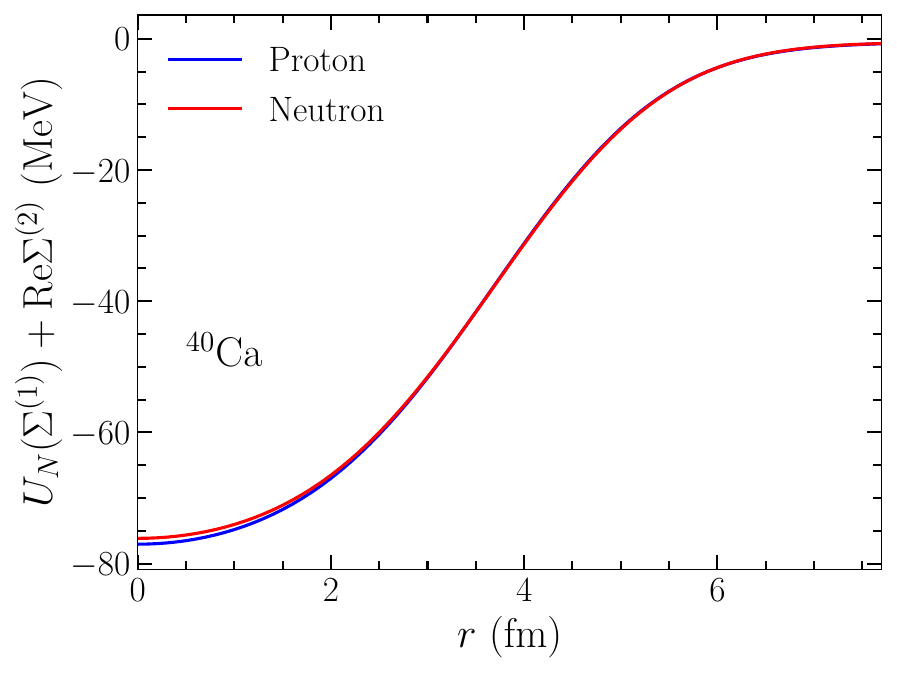}
   }\hfill
    \subfloat{
     \centering
     \includegraphics[width=0.46\linewidth]{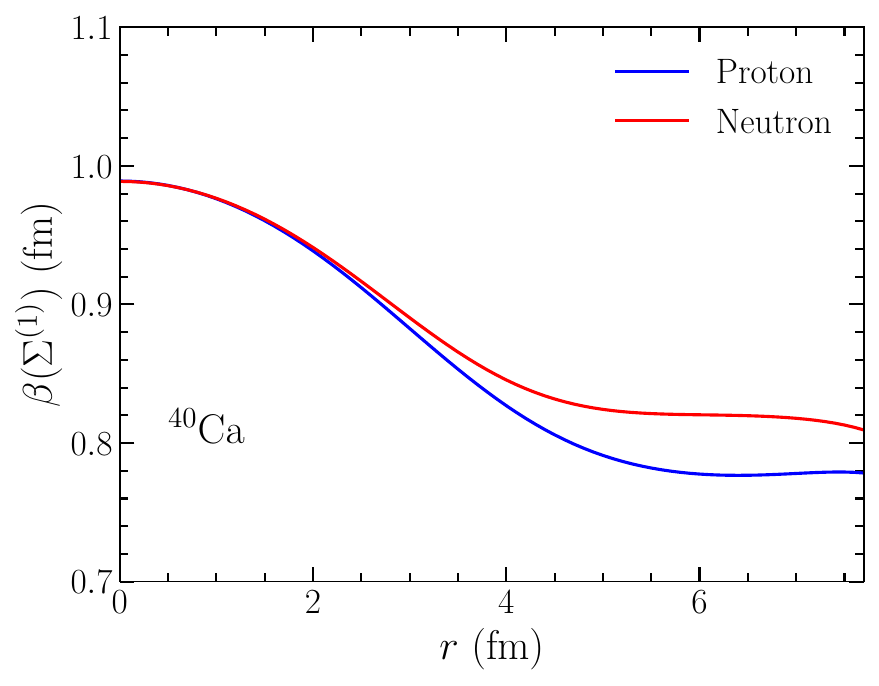}
   }
   \caption{Radial dependence of $U_N(\Sigma^{(1)})+\Sigma^{(2)}(q,E(q))$ and $\beta(\Sigma^{(1)})$ for $^{40}$Ca at a scattering energy of $E = 50$\,MeV.}
   \label{Fig:Ca401stOrderRDep}
\end{figure*}

\begin{figure*}[t]
    \subfloat{
     \centering
     \includegraphics[width=0.46\linewidth]{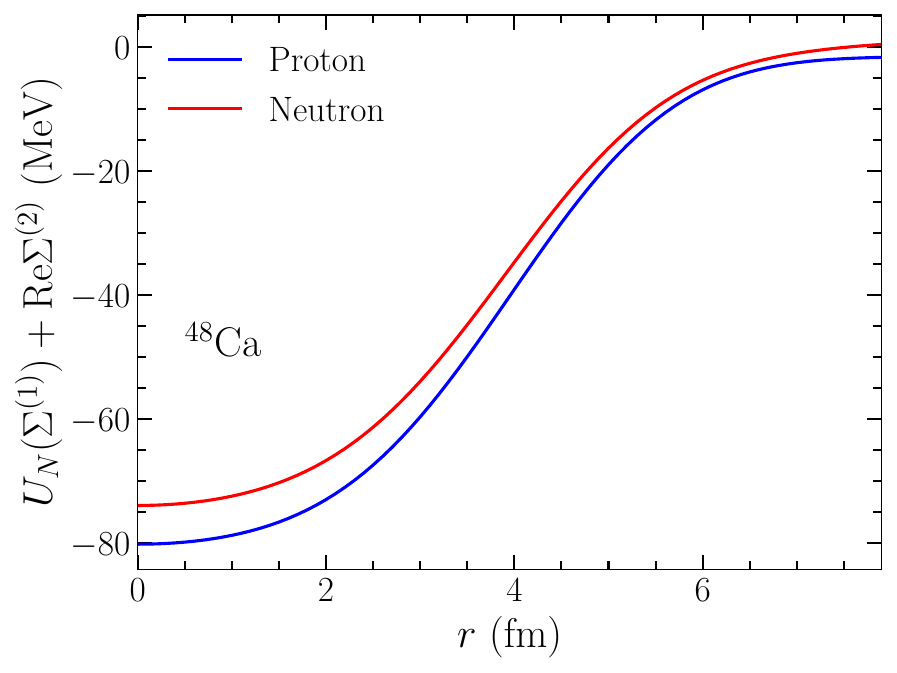}
     \label{Fig:Ca48U_NVsR1stOnly}
   }\hfill
    \subfloat{
     \centering
     \includegraphics[width=0.46\linewidth]{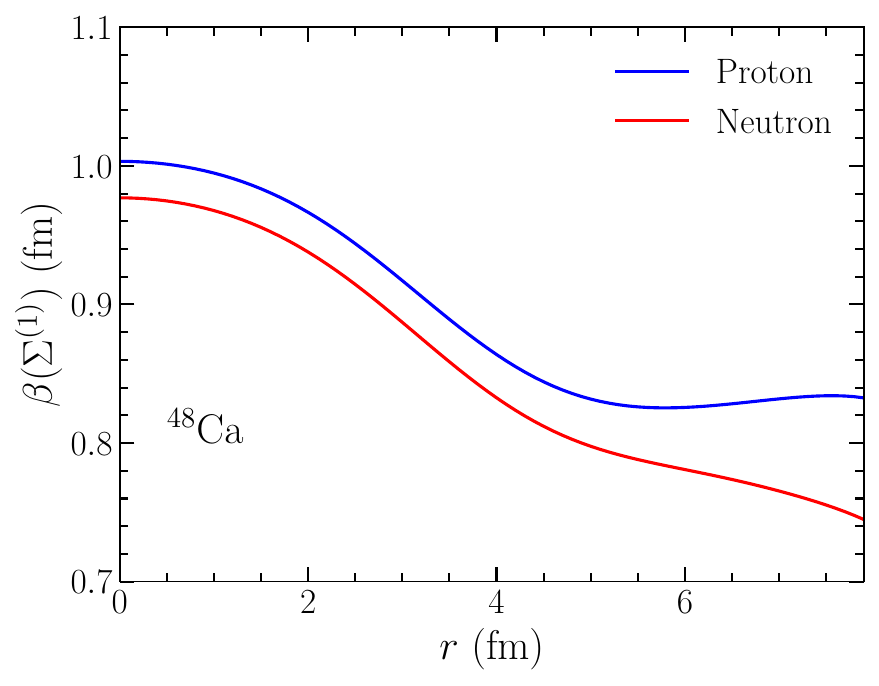}
     \label{Fig:Ca48BetaVsR1stOnly}
   }
   \caption{Radial dependence of $U_N(\Sigma^{(1)})+\Sigma^{(2)}(q,E(q))$ and $\beta(\Sigma^{(1)})$ for $^{48}$Ca at a scattering energy of $E = 50$\,MeV.}
   \label{Fig:Ca481stOrderRDep}
\end{figure*}

The $\beta$ parameter, which characterizes the range of influence of nonlocal effects, has been taken to be constant in density in previous phenomenological implementations \cite{Perey1962,Ripka1963,Titus2014}. In addition, these older phenomenological potentials contain a single $\beta$ parameter for both protons and neutrons.   Phenomenologically, in Ref.\ \cite{Perey1962}, the value of $\beta = 0.84\,\mathrm{fm}$ for 208Pb was obtained by fitting to low-energy neutron scattering data. Other investigations \cite{wyatt60,bowen63} similarly found $\beta = 0.82-0.87$\,fm for neutrons scattering on 208Pb. From Figures \ref{Fig:1stOrderRhoDep} and \ref{Fig:1stAnd2ndOrderRhoDep}, we see that the density-independence assumption is not globally supported by microscopic calculations, especially at first order in perturbation theory. However, the $\beta$ values obtained from chiral nuclear interactions, $\beta \sim 0.8-0.9$, are actually quite similar to phenomenological extractions \cite{Perey1962,wyatt60,bowen63}. Finally, we see from Figures \ref{Fig:1stOrderRhoDep} and \ref{Fig:1stAnd2ndOrderRhoDep} that there are significant differences between proton and neutron nonlocality strengths, which is also a notable feature of more modern nonlocal dispersive optical potentials  \cite{mahzoon17,atkinson20}. The above insights from microscopic many-body theory may enable future improvements in the phenomenological modeling of nonlocal potentials. 

To conclude this subsection, we return to the discussion of the isospin asymmetry dependence of the nonlocal parameters. Figure \ref{Fig:AsymmetryDep} shows the isospin asymmetry dependence of the $U_N$ and $\beta$ parameters fitted to the first-order (top) and first-order plus second-order (bottom) contributions to the self-energy in nuclear matter at density $n=n_0$. In all cases we see that the depth parameters $U_N$ exhibit an equal but opposite linear behavior for protons and neutrons at low values of $\delta_{np}$ that is consistent with the Lane approximation in Eq.\ \eqref{lane}. However, nonlinearities are seen to emerge at isospin asymmetries above $\delta \sim 0.2$. Interestingly, the nonlocality range parameter $\beta$ also exhibits a linear isovector character at low values of the isospin asymmetry. It can also be seen that the nonlocal strength and range parameters vary much more strongly with the isospin asymmetry when calculated from both the first- and second-order MBPT contributions compared to when they are extracted from the first-order contribution alone. Especially the higher-order $\delta_{np}^2$ terms, which are isoscalar in nature \cite{Holt2016}, are prominent in the isospin-asymmetry dependence of the $\beta$ parameter. These trends have previously been observed in the isospin-asymmetry dependence of the energy-dependent local optical potential depth $U_L(E)$ \cite{Holt2016}. From the right panels of Figure \ref{Fig:AsymmetryDep}, we see that in general $\beta$ does not vary strongly from the empirically derived value $\beta = 1\,\mathrm{fm}$ over the range of isospin asymmetries typically encountered in finite nuclei, except for the case of the neutron $\beta(\Sigma^{(1)})$ in highly neutron-rich matter. Therefore, it may be a reasonable approximation to use a single, averaged value of $\beta$ for scattering calculations, though this should be investigated in quantitative detail.

\subsection{Nonlocal optical potentials for calcium isotopes}
\label{SubSec:FirstVSecondOrder}

Starting from the Perey-Buck parameters $\beta$ and $U_N$, as well as the imaginary potential depth $W_L$, for various densities and proton fractions, we then employ the improved local-density approximation (ILDA) to obtain nucleon-nucleus nonlocal optical potentials. Here we focus on the nonlocal potentials for the calcium isotopic chain, from $^{40}$Ca to $^{60}$Ca. In this subsection we show the radial dependence of proton and neutron optical potential parameters for $^{40}$Ca and $^{48}$Ca, and we also study the dependence of the Woods-Saxon shape parameters on the atomic number across the calcium chain.

\begin{figure*}
    \subfloat{
     \centering
     \includegraphics[width=0.46\linewidth]{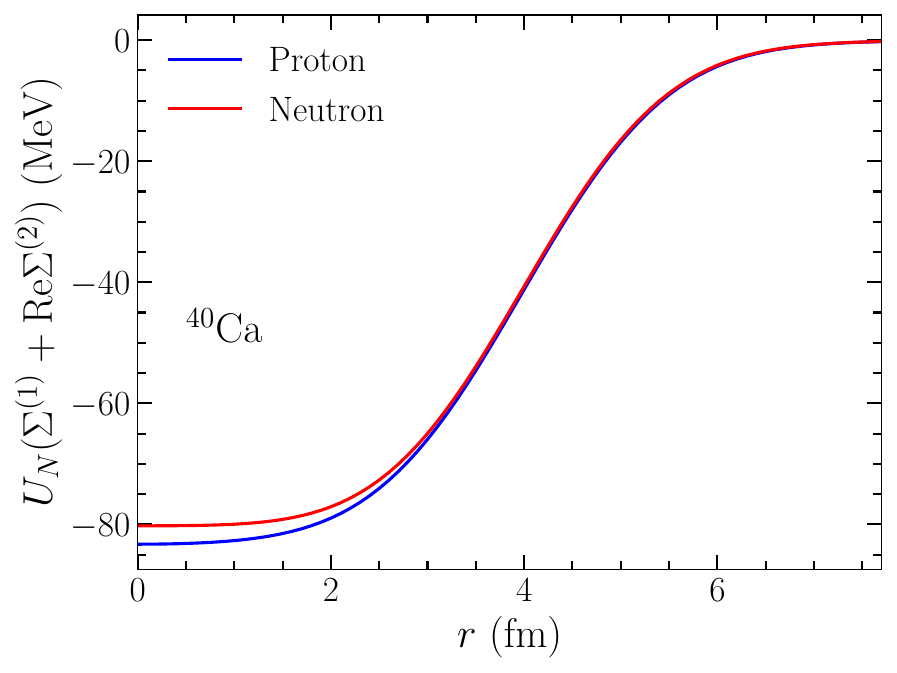}
   }\hfill
    \subfloat{
     \centering
     \includegraphics[width=0.46\linewidth]{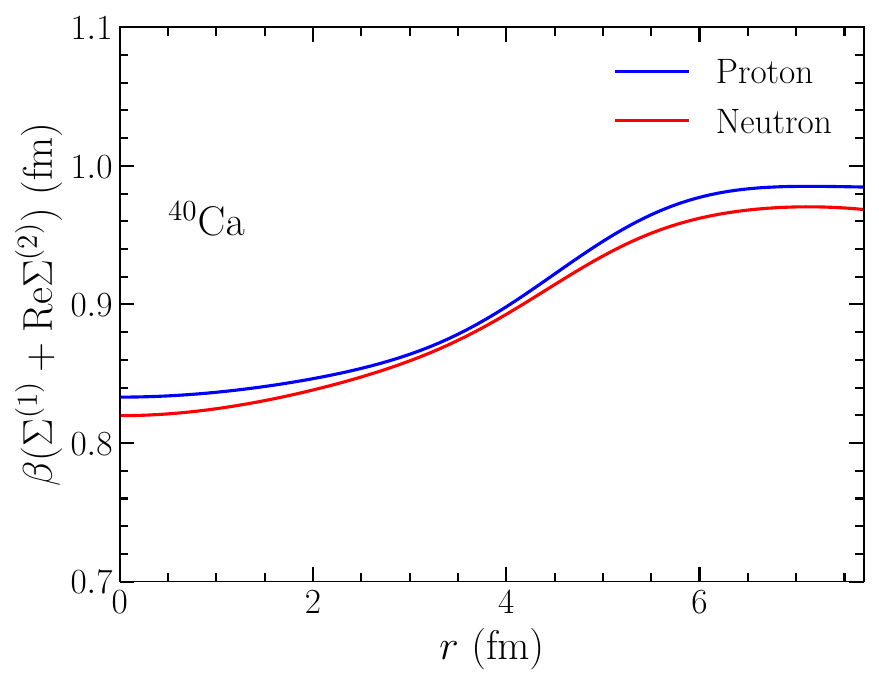}
   }
   \caption{Radial dependence of $U_N(\Sigma^{(1)}+\Sigma^{(2)})$ and $\beta(\Sigma^{(1)}+\Sigma^{(2)})$ for $^{40}$Ca.}
   \label{Fig:Ca401stAnd2ndRDep}
\end{figure*}

\begin{figure*}
    \subfloat{
     \centering
     \includegraphics[width=0.46\linewidth]{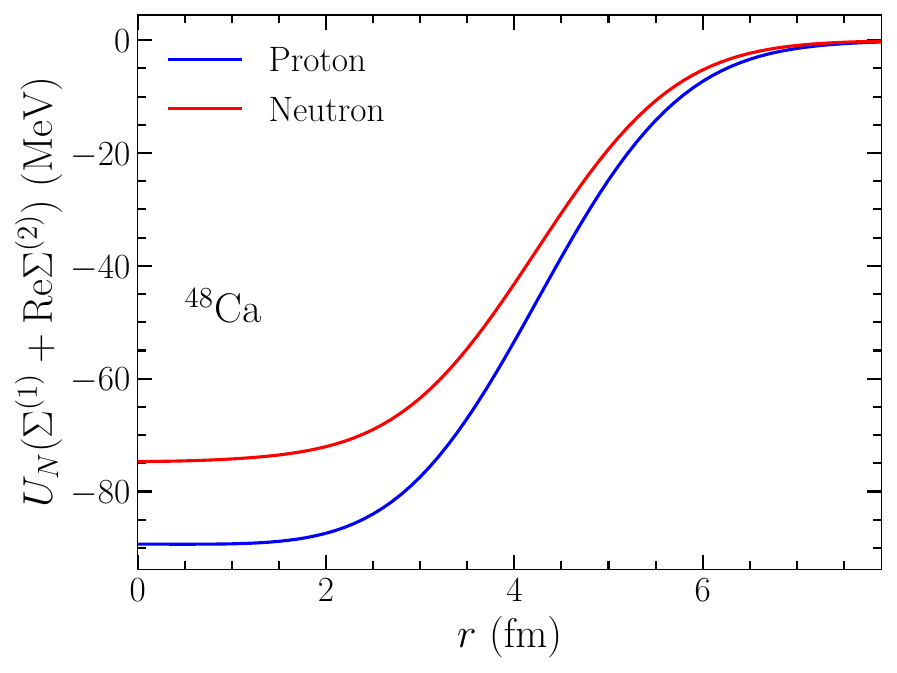}
   }\hfill
    \subfloat{
     \centering
     \includegraphics[width=0.46\linewidth]{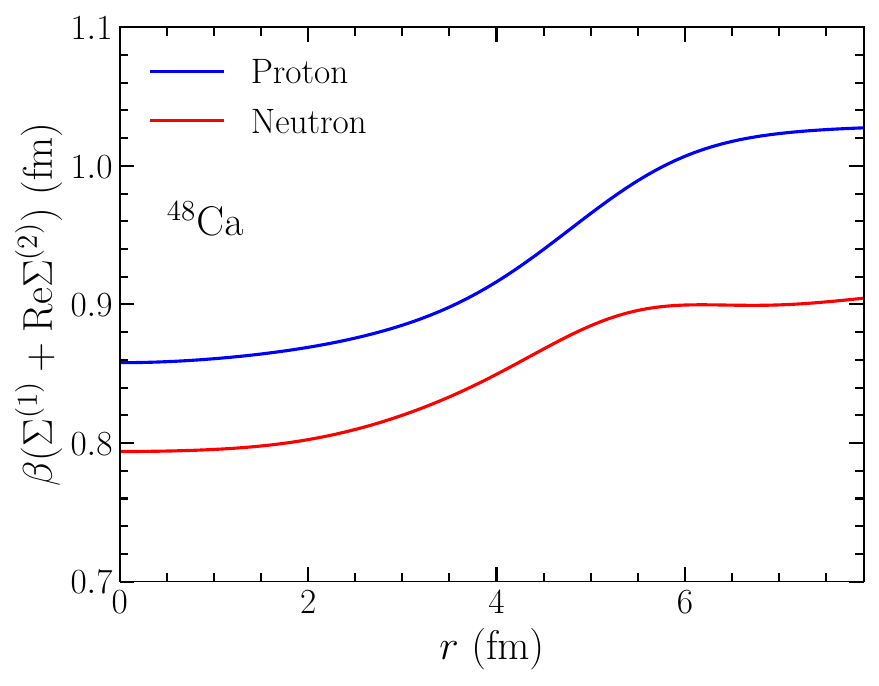}
   }
   \caption{Radial dependence of $U_N(\Sigma^{(1)}+\Sigma^{(2)})$ and $\beta(\Sigma^{(1)}+\Sigma^{(2)})$ for $^{48}$Ca.}
   \label{Fig:Ca481stAnd2ndRDep}
\end{figure*}

In the left panels of Figures \ref{Fig:Ca401stOrderRDep} and \ref{Fig:Ca481stOrderRDep}, we show the radial dependence of $U_N(\Sigma^{(1)})+\Sigma^{(2)}(q,E(q)=50\,\mathrm{MeV})$ for $^{40}$Ca and $^{48}$Ca, respectively. We note that this quantity contains an explicit dependence on the energy through the $\Sigma^{(2)}$ term. Essentially, we have assumed that the spatial nonlocality in the optical potential arises only from the first-order self-energy, which is strongly momentum dependent and has no explicit energy dependence. On the other hand, the second-order self-energy contribution is treated as local but energy dependent. This combination will provide a point of comparison for the fully nonlocal and energy-independent optical potentials for $^{40}$Ca and $^{48}$Ca, i.e.,  $U_N(\Sigma^{(1)}+\Sigma^{(2)})$, that we will discuss later. From the left panel of Figure \ref{Fig:Ca401stOrderRDep}, we see that for $^{40}$Ca the potential depth is approximately $80$\,MeV for both protons and neutrons. The average depth changes only slightly for protons and neutrons in $^{48}$Ca, where in the left panel of Figure \ref{Fig:Ca481stOrderRDep} we see a proton-neutron splitting of $\Delta U \simeq 7$\,MeV at the central density. These values are similar to the central depths $U_N = 70-75$\,MeV obtained in phenomenological studies of 208Pb \cite{Perey1962,wyatt60,bowen63}.

\begin{figure*}[t]
\centering
    \subfloat{
     \centering
     \includegraphics[width=0.32\textwidth]{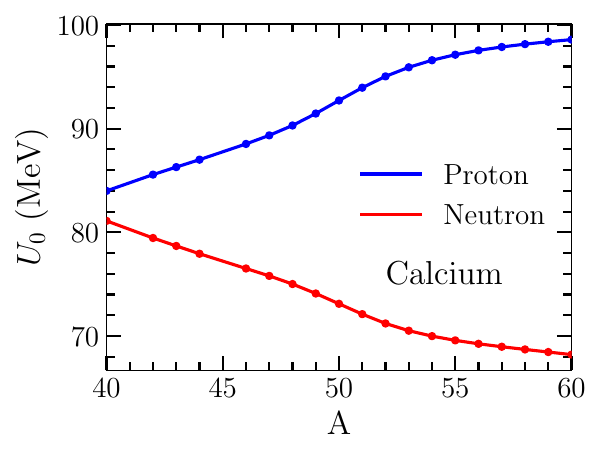}
   }
    \subfloat{
     \centering
     \includegraphics[width=0.32\textwidth]{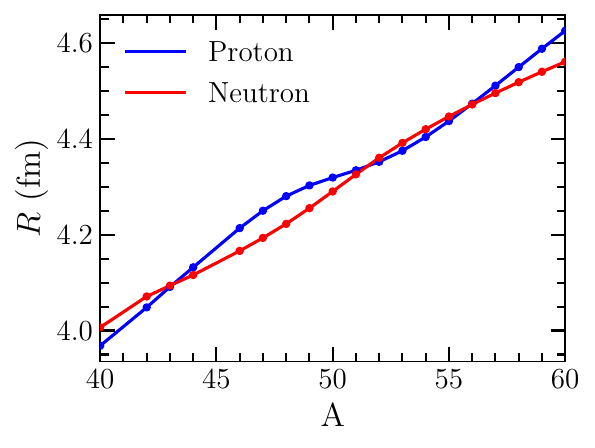}
   }
   \subfloat{
     \centering
     \includegraphics[width=0.32\textwidth]{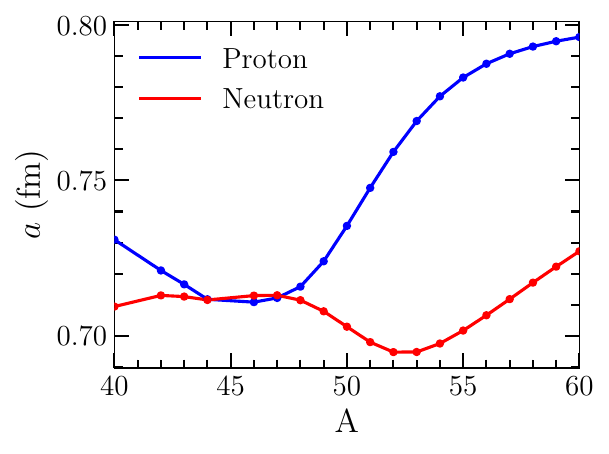}
   }
   \caption{Woods-Saxon parameters for proton and neutron nonlocal optical potential wells $U_N(r)$ as a function of the calcium isotope mass number $A$. (Left) radial dependence of the central potential depth $U_0$, (middle) radial dependence of the nucleus radius $R$, and (right) the radial dependence of the diffuseness $a$.}
   \label{Fig:1stAnd2ndUNWoodsSax}
\end{figure*}

In the right panels of Figures \ref{Fig:Ca401stOrderRDep} and \ref{Fig:Ca481stOrderRDep}, we show the radial dependence of $\beta(\Sigma^{(1)})$ for $^{40}$Ca and $^{48}$Ca, respectively. Since $\beta(\Sigma^{(1)})$ increases with the density, we see that for both $^{40}$Ca and $^{48}$Ca the proton and neutron nonlocality parameters decrease with the radial distance, from $\beta \simeq 1.0$\,fm at the center to $\beta \simeq 0.8$\,fm at the nuclear surface. We recall that phenomenological values of $\beta$ lie in the range $\beta \simeq 0.8-0.9$, which is similar to the average value obtained in our analysis. In addition, similar microscopic studies of the Perey-Buck nonlocality have found values for $\beta$ near our average value for $^{40}$Ca, with $\beta \approx 0.85 \mathrm{fm}$ \cite{Arellano2022}. In $^{48}$Ca there are small but noticeable differences in the value of $\beta$ for protons and neutrons, with $\beta_p - \beta_n \simeq 0.03$\,fm, which is roughly constant over the length of the nucleus.

In the left-hand panels of Figures \ref{Fig:Ca401stAnd2ndRDep} and \ref{Fig:Ca481stAnd2ndRDep}, we show the radial dependence of $U_N(\Sigma^{(1)}+\Sigma^{(2)})$ for $^{40}$Ca and $^{48}$Ca, respectively. In the case of $^{40}$Ca, the nonlocal potential strength at the center of the nucleus is approximately 10\% stronger than $U_N(\Sigma^{(1)})+\Sigma^{(2)}(q,E(q)=50\,\mathrm{MeV})$ for both proton and neutron projectiles. In the case of $^{48}$Ca, the large isovector optical potential in the $U_N(\Sigma^{(1)}+\Sigma^{(2)})$ approximation leads to a larger splitting of the proton and neutron central well depths compared to the results of Figure \ref{Fig:Ca481stOrderRDep} for the $U_N(\Sigma^{(1)})+\Sigma^{(2)}(q,E(q)=50\,\mathrm{MeV})$ approximation. This leads to neutron optical potential depths that are very similar in both treatments of the nonlocality, but proton potential depths that differ by approximately 20\%. Since $U_N(\Sigma^{(1)}+\Sigma^{(2)})$ lacks the energy dependence present in the $U_N(\Sigma^{(1)})+\Sigma^{(2)}(q,E(q))$ approximation, the comparison of these optical potentials at the specific energy $E=50$\,MeV is only for orientation purposes.

In the right-hand panels of Figures \ref{Fig:Ca401stAnd2ndRDep} and \ref{Fig:Ca481stAnd2ndRDep}, we show the radial dependence of $\beta(\Sigma^{(1)}+\Sigma^{(2)})$ for $^{40}$Ca and $^{48}$Ca, respectively. It can be seen that in $^{48}$Ca the difference in $\beta$ values for protons and neutrons is much larger when calculated from the combined first- and second-order contributions to the self-energy. The radial dependence of $\beta(\Sigma^{(1)}+\Sigma^{(2)})$ also shows the opposite behavior compared to that of $\beta(\Sigma^{(1)})$, increasing to larger values as the nucleon density decreases near the surface of the nucleus. 
However, $\beta$ still does not change much throughout the nuclear volume, with a maximum difference of only $\sim$10\%.
\subsection{Woods-Saxon parameterizations and their isotopic dependence}

A visual inspection of the above figures suggests that the potential depth $U_N(\Sigma^{(1)}+\Sigma^{(2)})$ matches the typical Woods-Saxon form
\begin{equation}
    U_N(r) = -\frac{U_0}{1+\mathrm{exp}((r-R_U)/a_U)}
    \label{StandardWoodsSaxon}
\end{equation}
used to describe the radial dependence of phenomenological optical potentials, and in fact a fit to this form can be easily made. This may help simplify the implementation of the new potentials in existing scattering codes. The parameter $\beta$ also appears to have a similar behavior, though with a modified Woods-Saxon form:
\begin{equation}
   \beta(r) = \frac{\beta_0}{1 + \mathrm{exp}((r - R_\beta) / a_\beta)} + \Delta_\beta,
\end{equation}
where $\Delta_\beta$ is an added offset parameter that accounts for the non-zero asymptotic value of $\beta$. Detailed fits found that the radial dependence of $\beta$ cannot be completely captured by a Woods-Saxon form, but given how little the value of $\beta$ varies within the nucleus, a Woods-Saxon parameterization was found to be reasonable.

In Figure \ref{Fig:1stAnd2ndUNWoodsSax} we show the isotopic dependence of the fitted Woods-Saxon parameters across the Ca isotopic chain as a function of the mass number $A$. It can be seen that the dependence of these parameters on the mass number of the target nucleus is very smooth, suggesting that it will be possible in the future to construct a microscopic nonlocal optical potential across the chart of nuclides, similar to what has already been performed for microscopic local optical potentials \cite{Whitehead2021}. 

In addition, Figure \ref{Fig:1stAnd2ndUNWoodsSax} also demonstrates an inflection point in all Woods-Saxon parameters around $A = 48$; this is likely related to shell effects and the double-shell closure at $^{48}$Ca. Therefore, the addition of neutrons on top of the doubly-filled shells introduces qualitatively new behaviors. As expected, the proton and neutron nonlocal optical potential depths $U_0$ diverge symmetrically starting at $^{40}$Ca, where the initial small difference in optical potential depths at $r=0$ is due to the neutron excess created by Coulomb repulsion. The dependence of the nuclear radius parameter $R$ on the mass number is shown in the middle panel of Figure \ref{Fig:1stAnd2ndUNWoodsSax}. Interestingly, for $^{40}$Ca the proton radius parameter is smaller than the neutron radius parameter, but the larger proton diffuseness parameter for $^{40}$Ca shown in the right panel of Figure \ref{Fig:1stAnd2ndUNWoodsSax} gives rise to a more extended proton distribution in $^{40}$Ca compared to the neutron distribution. 

\begin{figure*}
    \subfloat{
     \centering
     \includegraphics[width=0.46\linewidth]{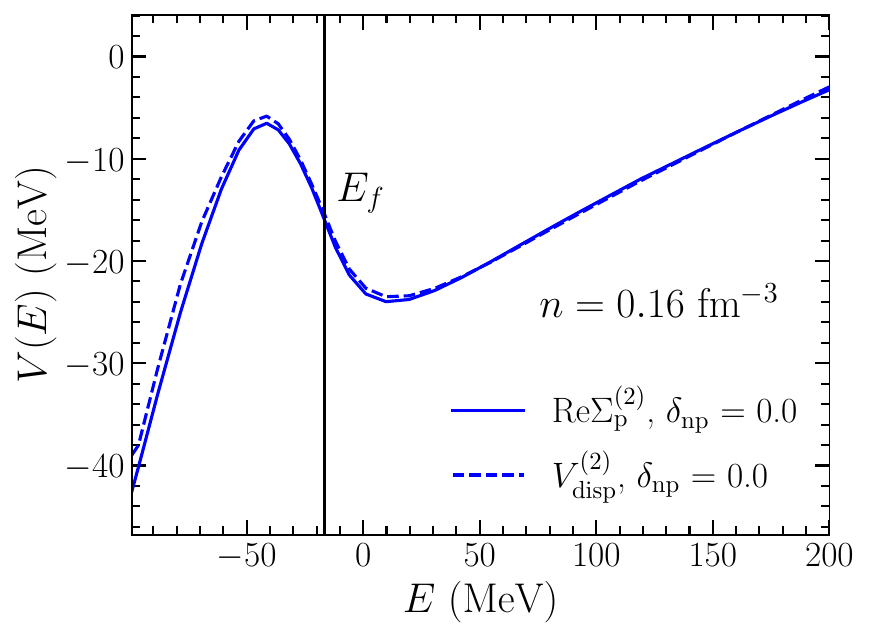}
   }\hfill
   \subfloat{
     \centering
     \includegraphics[width=0.46\linewidth]{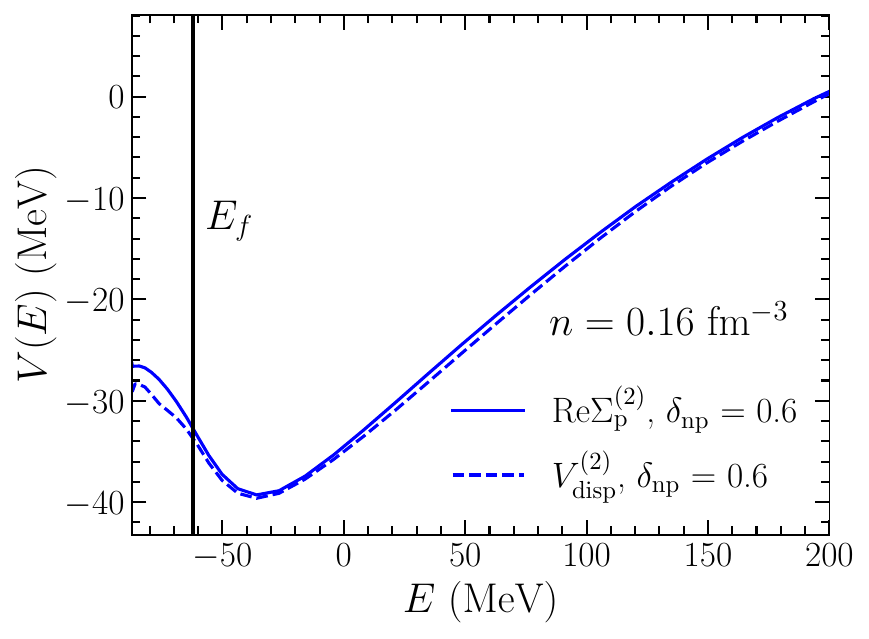}
   }
   \caption{The second-order real proton self-energy in nuclear matter at saturation density and two isospin asymmetries: $\delta_{np} = 0$ (left) and $\delta_{np} = 0.6$ (right). The solid lines denote the result of direct calculation of the second-order perturbation theory diagrams, while the dashed lines show the results obtained from the dispersion relation in Eq.\ \eqref{dr1}.}
   \label{Fig:NMVDispersiveComparison1}
\end{figure*}

\subsection{Optical potential dispersion relations}
\label{Dispersivity}

As we discussed at the end of Section \ref{SubSec:PBNonlocalResults}, the energy dependence of the imaginary potential $W_{L}(E)$ cannot be fitted using the Perey-Buck ansatz. This introduces a fundamental physical inconsistency when constructing a complete optical model description. While the microscopic real self-energy up to second order appears to be well-parameterized by an energy-independent nonlocal potential, pairing it with a local, energy-dependent imaginary part would violate the structure required by causality and the Kramers-Kronig dispersion relation:
\begin{equation}
\text{Re}\Sigma(q,E) = \Sigma_{\text{HF}}(q) - \frac{1}{\pi} \mathcal{P} \int_{-\infty}^{\infty} \frac{\text{Im}\Sigma(q,E')}{E' - E} dE',
\label{dr1}
\end{equation}
where $\Sigma_{\text{HF}}(q)$ is the energy-independent Hartree-Fock component, and we note that the dispersion integral extends over both positive energies (scattering) and negative energies (bound-states). In general, the evaluation of the dispersion integral on the right-hand side of Eq.\ \eqref{dr1} will lead to a residual dependence on the energy $E$.
Enforcing an energy-independent nonlocal representation exclusively on the real potential while leaving the imaginary part entirely local and energy-dependent would therefore break this relationship.

We note that purely phenomenological optical potentials often do not obey a dispersion relation, since their real and imaginary parts are in general constructed independently from one another, with the intent of reproducing experimental data. In addition, optical potentials constructed solely to describe positive-energy scattering observables would not cover the relevant negative-energy domain of the dispersion relation. By contrast, dispersive optical models take advantage of this mathematical property to encode resonance and absorption properties of a particular nucleus that can be measured precisely and are strongly correlated with the imaginary part of the optical potential \cite{Mahzoon2014}.

In our work, the nucleon self-energy in nuclear matter is expected to be exactly dispersive by construction, as it is derived without approximation from the nucleon propagator. This can be seen in Figure \ref{Fig:NMVDispersiveComparison1}, where we show the real part of the proton self-energy in nuclear matter at saturation density and at two different choices of the isospin asymmetry: $\delta_{np} = 0$ (left panel) and $\delta_{np} = 0.6$ (right panel). Although small numerical discrepancies appear due to the finite energy range over which we compute the nucleon self-energy, we find that overall the real self-energy calculated explicitly at second order in perturbation theory (solid lines) agrees well with that obtained by integrating the off-shell second-order imaginary contributions using the dispersion integral of Eq.\ \eqref{dr1}.

We note that the validity of Eq.\ \ref{dr1} requires that the momentum be fixed as the imaginary self-energy is integrated with respect to the energy along the real axis. It is therefore clear that the on-shell approximation in Eq.\ \eqref{Eq:SelfConsistentE} leads to a violation of the dispersion relation; rather than keeping the momentum fixed, the momentum varies along its on-shell value over the intermediate $E'$ integration:
\begin{equation}
V(q(E), E) = V_{\text{HF}}(q(E))-\frac{1}{\pi} \mathcal{P} \int_{-\infty}^{\infty} \frac{W(q(E'), E')}{E' - E} dE'.
\end{equation}

\begin{figure*}
    \subfloat{
     \centering
     \includegraphics[width=0.46\linewidth]{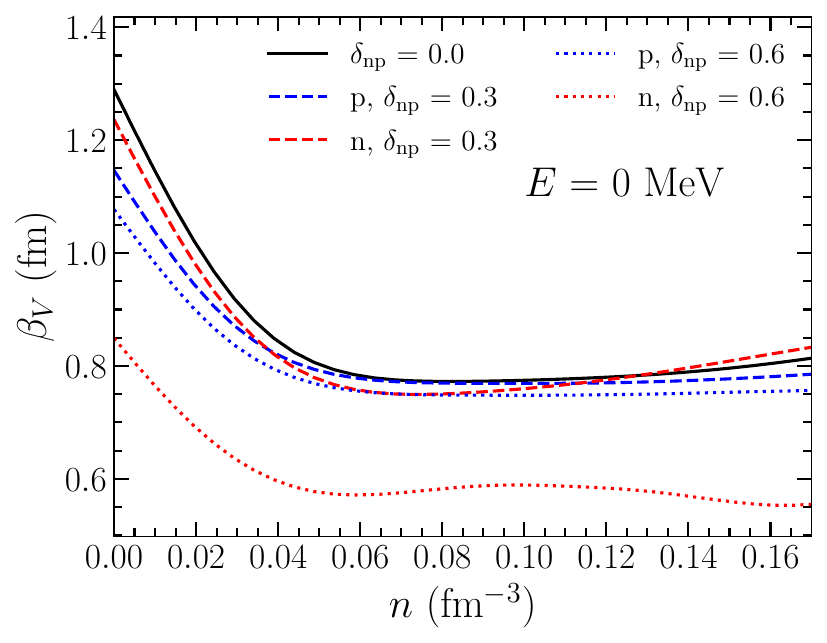}
   }\hfill
   \subfloat{
     \centering
     \includegraphics[width=0.46\linewidth]{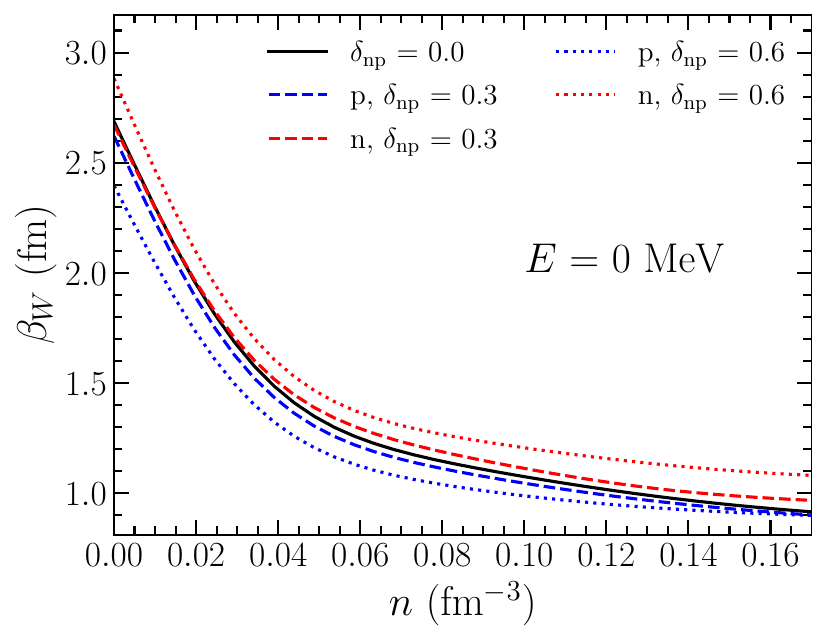}
   } \protect\\
   
   \subfloat{
     \centering
     \includegraphics[width=0.46\linewidth]{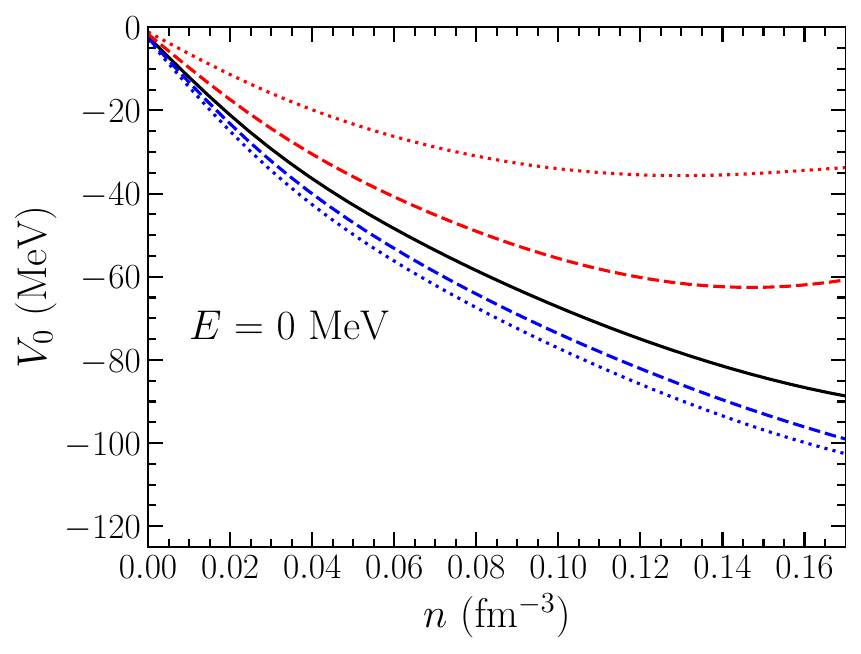}
   }\hfill
   \subfloat{
     \centering
     \includegraphics[width=0.46\linewidth]{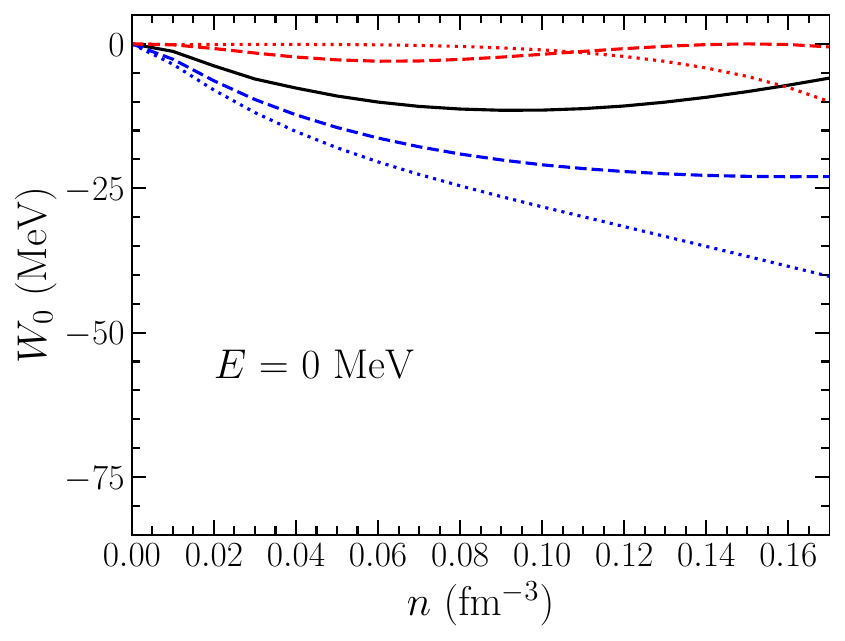}
   }
   \caption{Density-dependent beta parameters and nonlocal potential depths extracted directly from the real (left) and imaginary (right) self-energy.} 
   \label{Fig:NMVDispersiveComparison2}
\end{figure*}

In order to construct a dispersive optical potential, it is therefore necessary to Fourier transform from momentum space to coordinate space:
\begin{equation}
\text{Re}\Sigma(\vec{r}, \vec{r\,}'; E) = \Sigma_{\text{HF}}(\vec{r}, \vec{r\,}') - \frac{1}{\pi} \mathcal{P} \int_{-\infty}^{\infty} \frac{\text{Im}\Sigma(\vec{r}, \vec{r\,}'; E')}{E' - E} dE',
\end{equation}
which fully encodes the constant-momentum requirement into the independent (and nonlocal) spatial variables. For infinite nuclear matter, translation invariance implies $\Sigma(\vec{r}, \vec{r\,}'; E) = \Sigma(|\vec{r} - \vec{r\,}'|; E)$. Using the local-density approximation it is then possible to perform an identical calculation at an arbitrary radial position in a finite nucleus by simply treating that point as uniform matter with conditions determined by the nucleon density distribution at that point. In this way one obtains a dispersive potential that exactly matches the real self-energy calculated at that same point.

We have found that fitting a simple Gaussian form to the momentum dependence of ${\rm Re}(\Sigma^{(1)}+\Sigma^{(2)})$ and then Fourier transforming to position space leads to nonlocal real depths and ranges consistent with what we have found above in Section \ref{SubSec:DensityDep}. In the left panels of Figure \ref{Fig:NMVDispersiveComparison2} we show the density dependence of the extracted nonlocality parameters for the real self-energy at $E=0$. Compared to the values of the nonlocal depth and range parameters found from fitting to the Perey-Buck form (Figure \ref{Fig:1stAnd2ndOrderRhoDep}), we see that those obtained using a direct Gaussian fit of the off-shell self-energy differ by less than 10\%. These differences can be explained by noting that the Gaussian fit of the self-energy momentum dependence described here leaves unchanged the energy dependence of the self-energy, whereas the Perey-Buck fitting described in Section \ref{SubSec:DensityDep} absorbs all of the on-shell energy dependence into a Gaussian spatial nonlocality. In addition, we show in the right panels of Figure \ref{Fig:NMVDispersiveComparison2} the nonlocality parameters for ${\rm Im}\Sigma^{(2)}$ at energy $E=0$ obtained by directly fitting the momentum dependence to a Gaussian form. In the future, a dispersive nonlocal optical potential may be constructed by systematically extending this method across all energies.

\section{Conclusion}
\label{Sec:Conclusion}
We have investigated the Perey-Buck parameterization for proton and neutron optical potential nonlocalities in infinite nuclear matter starting from chiral effective field theory two-body and three-body forces. We have found that nearly energy-independent optical potentials can be obtained by including just the first-order perturbative contribution to the self-energy or including both first- and second-order contributions to the self-energy, thereby validating the Perey-Buck ansatz for microscopic optical potentials. We find that the range of scattering energies for which this is valid is approximately $0<E<100$\,MeV, beyond which the Perey-Buck parameterization breaks down. Although the proton and neutron nonlocal potential depths $U_N$ and range parameters $\beta$ in nuclear matter at saturation density generally lie within the range obtained from previous phenomenological fits, we also find systematic dependencies of these parameters on the density and isospin asymmetry, which are not typically included in phenomenological implementations.

Starting from the proton and neutron self-energies in nuclear matter at varying density and isospin asymmetry, we construct nonlocal nucleon-nucleus optical potentials for calcium isotopes employing the improved local density approximation. We observe significant splittings of the proton and neutron nonlocal potential depth parameter $U_N$ and nonlocality range parameter $\beta$. In particular, the nonlocality range parameter for protons is larger than that for neutrons in $^{48}$Ca. We also demonstrate that $\beta$ varies significantly with radial distance. Finally, we have found a smooth dependence of the Woods-Saxon shape parameters on the mass number $A$ in the Calcium isotopic chain. 

In the future, we plan to compare results for elastic scattering and transfer reactions computed using the derived nonlocal optical potentials to those from the WLH energy-dependent local optical potential constructed from the same underlying two-body and three-body chiral nuclear forces. This will provide insight into theoretical uncertainties associated with the on-shell approximation for the energy- and momentum-dependent self-energy. We also plan to construct a modified version of the WLH microscopic global optical potential with the real volume term containing spatial nonlocalities inherent to the optical potential while avoiding assumptions that might limit the potential to cases of elastic scattering.

\section*{Acknowledgements}
This work was supported in part by the US Department of Energy, National Nuclear Security Administration under Awards DE-NA0003841 and DE-NA0004150. Additional support was provided by the National Science Foundation under Grant No.\ PHY2209318. Portions of this research
were conducted with the advanced computing resources
provided by Texas A\&M High Performance Research
Computing.

\section*{Data Availability}
The data that support the findings of this article are publicly available \cite{stahulak_2025_17102397}.
\bibliographystyle{apsrev4-1}
\bibliography{export}

@article{mahzoon17,
  title = {Neutron Skin Thickness of $^{48}\mathrm{Ca}$ from a Nonlocal Dispersive Optical-Model Analysis},
  author = {Mahzoon, M. H. and Atkinson, M. C. and Charity, R. J. and Dickhoff, W. H.},
  journal = {Phys. Rev. Lett.},
  volume = {119},
  issue = {22},
  pages = {222503},
  numpages = {5},
  year = {2017},
  month = {Nov},
  publisher = {American Physical Society},
  doi = {10.1103/PhysRevLett.119.222503},
  url = {https://link.aps.org/doi/10.1103/PhysRevLett.119.222503}
}

@article{atkinson20,
  title = {Dispersive optical model analysis of $^{208}\mathrm{Pb}$ generating a neutron-skin prediction beyond the mean field},
  author = {Atkinson, M. C. and Mahzoon, M. H. and Keim, M. A. and Bordelon, B. A. and Pruitt, C. D. and Charity, R. J. and Dickhoff, W. H.},
  journal = {Phys. Rev. C},
  volume = {101},
  issue = {4},
  pages = {044303},
  numpages = {15},
  year = {2020},
  month = {Apr},
  publisher = {American Physical Society},
  doi = {10.1103/PhysRevC.101.044303},
  url = {https://link.aps.org/doi/10.1103/PhysRevC.101.044303}
}

@article{epelbaum09rmp,
  title = {Modern theory of nuclear forces},
  author = {Epelbaum, E. and Hammer, H.-W. and Mei\ss{}ner, Ulf-G.},
  journal = {Rev. Mod. Phys.},
  volume = {81},
  pages = {1773},
  year = {2009}
}

@article{machleidt11,
 author = {R. Machleidt and D. R. Entem},
 title = {Chiral effective field theory and nuclear forces},
 journal = {Phys. Rept.}, 
 volume = {503},
 pages = {1},
 year = {2011}
 }

@article{Holt:2011eg,
title = {Second-order quasiparticle interaction in nuclear matter with chiral two-nucleon interactions},
journal = {Nucl. Phys. A},
volume = {870-871},
pages = {1-22},
year = {2011},
issn = {0375-9474},
doi = {https://doi.org/10.1016/j.nuclphysa.2011.09.006},
url = {https://www.sciencedirect.com/science/article/pii/S0375947411005999},
author = {J.W. Holt and N. Kaiser and W. Weise}
}

@article{Holt:2011yj,
title = {Quasiparticle interaction in nuclear matter with chiral three-nucleon forces},
journal = {Nucl. Phys. A},
volume = {876},
pages = {61-76},
year = {2012},
issn = {0375-9474},
doi = {https://doi.org/10.1016/j.nuclphysa.2011.12.001},
url = {https://www.sciencedirect.com/science/article/pii/S0375947411006701},
author = {J.W. Holt and N. Kaiser and W. Weise}
}

@article{Arellano2022,
   author = {H. F. Arellano and G. Blanchon},
   doi = {10.1140/epja/s10050-022-00777-9},
   pages = {119},
   issue = {7},
   journal = {EPJ A},
   month = {6},
   title = {On the separability of microscopic optical model potentials and emerging bell-shape Perey-Buck nonlocality},
   volume = {58},
   url = {http://arxiv.org/abs/2206.09461 http://dx.doi.org/10.1140/epja/s10050-022-00777-9},
   year = {2022}
}

@article{Arellano2019,
   author = {H. F. Arellano and G. Blanchon},
   doi = {10.1016/j.physletb.2018.12.004},
   issn = {03702693},
   journal = {Phys. Lett. B},
   month = {2},
   pages = {256-261},
   publisher = {Elsevier B.V.},
   title = {Exact scattering waves off nonlocal potentials under Coulomb interaction within Schrödinger's integro-differential equation},
   volume = {789},
   year = {2019}
}

@article{Arellano2018,
   author = {H. F. Arellano and G. Blanchon},
   doi = {10.1103/PhysRevC.98.054616},
   pages = {054616},
   issn = {24699993},
   issue = {5},
   journal = {Phys. Rev. C},
   month = {11},
   publisher = {American Physical Society},
   title = {Irreducible nonlocality of optical model potentials based on realistic NN interactions},
   volume = {98},
   year = {2018}
}

@article{Balantekin2014,
   author = {A. B. Balantekin and J. Carlson and D. J. Dean and G. M. Fuller and R. J. Furnstahl and M. Hjorth-Jensen and R. V.F. Janssens and Bao An Li and W. Nazarewicz and F. M. Nunes and W. E. Ormand and S. Reddy and B. M. Sherrill},
   doi = {10.1142/S0217732314300109},
   pages = {1430010},
   issn = {02177323},
   issue = {11},
   journal = {Mod. Phys. Lett. A},
   month = {4},
   publisher = {World Scientific Publishing Co. Pte Ltd},
   title = {Nuclear theory and science of the facility for rare isotope beams},
   volume = {29},
   year = {2014}
}

@article{Bauge1998,
   author = {E Bauge and J P Delaroche and M Girod},
   isbn = {22.0298.523760.2},
   issue = {2},
   journal = {Phys. Rev. C},
   month = {8},
   pages = {1118-1145},
   title = {Semimicroscopic nucleon-nucleus spherical optical model for nuclei with A40 at energies up to 200 MeV},
   volume = {58},
   year = {1998}
}

@article{Bogner2009,
   author = {S. K. Bogner and R. J. Furnstahl and L. Platter},
   doi = {10.1140/epja/i2008-10695-1},
   issn = {14346001},
   issue = {2},
   journal = {EPJ A},
   month = {2},
   pages = {219-241},
   title = {Density matrix expansion for low-momentum interactions},
   volume = {39},
   year = {2009}
}

@article{Bonaccorso2018,
   author = {Angela Bonaccorso},
   doi = {10.1016/j.ppnp.2018.01.005},
   issn = {01466410},
   journal = {Prog. Part. Nucl. Phys.},
   month = {7},
   pages = {1-54},
   publisher = {Elsevier B.V.},
   title = {Direct reaction theories for exotic nuclei: An introduction via semi-classical methods},
   volume = {101},
   year = {2018}
}

@article{Brieva1977,
   author = {F. A. Brieva and J. R. Rook},
   issue = {2},
   journal = {Nucl. Phys. A},
   month = {3},
   pages = {317-341},
   title = {NUCLEON-NUCLEUS OPTICAL MODEL POTENTIAL (II). Finite Nuclei},
   volume = {291},
   year = {1977}
}

@article{Deltuva2009,
   author = {A. Deltuva},
   doi = {10.1103/PhysRevC.79.021602},
   pages = {021602},
   issn = {1089490X},
   issue = {2},
   journal = {Phys. Rev. C},
   month = {2},
   publisher = {American Physical Society},
   title = {Three-body direct nuclear reactions: Nonlocal optical potential},
   volume = {79},
   year = {2009}
}

@article{Dickhoff2019,
   author = {W. H. Dickhoff and R. J. Charity},
   doi = {10.1016/j.ppnp.2018.11.002},
   issn = {01466410},
   journal = {Prog. Part. Nucl. Phys.},
   month = {3},
   pages = {252-299},
   publisher = {Elsevier B.V.},
   title = {Recent developments for the optical model of nuclei},
   volume = {105},
   year = {2019}
}

@article{Giannini1976,
   author = {M M Giannini and G Ricco},
   journal = {Ann. Phys.},
   month = {1},
   pages = {458-492},
   title = {An Energy-independent Nonlocal Potential Model for Bound and Scattering States},
   volume = {102},
   year = {1976}
}

@article{Goriely2007,
   author = {S. Goriely and J. P. Delaroche},
   doi = {10.1016/j.physletb.2007.07.046},
   issn = {03702693},
   issue = {2-4},
   journal = {Phys. Lett. B},
   month = {9},
   pages = {178-183},
   title = {The isovector imaginary neutron potential: A key ingredient for the r-process nucleosynthesis},
   volume = {653},
   year = {2007}
}

@article{Hebborn102022,
   author = {C. Hebborn and F. M. Nunes and G. Potel and W. H. Dickhoff and J. W. Holt and M. C. Atkinson and R. B. Baker and C. Barbieri and G. Blanchon and M. Burrows and R. Capote and P. Danielewicz and M. Dupuis and Ch. Elster and J. E. Escher and L. Hlophe and A. Idini and H. Jayatissa and B. P. Kay and K. Kravvaris and J. J. Manfredi and A. Mercenne and B. Morillon and G. Perdikakis and C. D. Pruitt and G. H. Sargsyan and I. J. Thompson and M. Vorabbi and T. R. Whitehead},
   doi = {10.1088/1361-6471/acc348},
   issue = {6},
   journal = {J. Phys. G: Nucl. Part. Phys.},
   month = {10},
   pages = {060501},
   title = {Optical potentials for the rare-isotope beam era},
   volume = {50},
   url = {http://arxiv.org/abs/2210.07293 http://dx.doi.org/10.1088/1361-6471/acc348},
   year = {2022}
}

@article{Hebborn062022,
   author = {Chloë Hebborn and Gregory Potel},
   doi = {10.1103/PhysRevC.107.014607},
   issue = {1},
   journal = {Phys. Rev. C},
   month = {6},
   pages = {014607},
   title = {Green's Function Knockout Formalism},
   volume = {107},
   url = {http://arxiv.org/abs/2206.09948 http://dx.doi.org/10.1103/PhysRevC.107.014607},
   year = {2022}
}

@article{Hebborn092021,
   author = {C. Hebborn and F. M. Nunes},
   doi = {10.1103/PhysRevC.104.034624},
   pages = {034624},
   issn = {24699993},
   issue = {3},
   journal = {Phys. Rev. C},
   month = {9},
   publisher = {American Physical Society},
   title = {Considering nonlocality in the optical potentials within eikonal models},
   volume = {104},
   year = {2021}
}

@article{Hebborn2020,
   author = {C. Hebborn and P. Capel},
   doi = {10.1088/1742-6596/1643/1/012088},
   issn = {1742-6588},
   issue = {1},
   journal = {J. Phys. Conf. Ser.},
   month = {12},
   pages = {012088},
   publisher = {IOP Publishing},
   title = {Sensitivity of one-neutron knockout of halo nuclei to their nuclear structure},
   volume = {1643},
   year = {2020}
}

@article{Holt2013,
   author = {J. W. Holt and N. Kaiser and G. A. Miller and W. Weise},
   doi = {10.1103/PhysRevC.88.024614},
   pages = {024614},
   issn = {1089490X},
   issue = {2},
   journal = {Phys. Rev. C},
   month = {8},
   publisher = {American Physical Society},
   title = {Microscopic optical potential from chiral nuclear forces},
   volume = {88},
   year = {2013}
}

@article{Holt2016,
   author = {J. W. Holt and N. Kaiser and G. A. Miller},
   doi = {10.1103/PhysRevC.93.064603},
   pages = {064603},
   issn = {24699993},
   issue = {6},
   journal = {Phys. Rev. C},
   month = {6},
   publisher = {American Physical Society},
   title = {Microscopic optical potential for exotic isotopes from chiral effective field theory},
   volume = {93},
   year = {2016}
}

@article{wyatt60,
  title = {Nonlocal Optical Model for Nucleon-Nuclear Interactions},
  author = {Wyatt, P. J. and Wills, J. G. and Green, A. E. S.},
  journal = {Phys. Rev.},
  volume = {119},
  issue = {3},
  pages = {1031--1042},
  numpages = {0},
  year = {1960},
  month = {Aug},
  publisher = {American Physical Society},
  doi = {10.1103/PhysRev.119.1031},
  url = {https://link.aps.org/doi/10.1103/PhysRev.119.1031}
}

@article{bowen63,
title = {Small angle elastic scattering of neutrons by uranium at energies between 18 MeV and 120 MeV},
journal = {Nucl. Phys.},
volume = {40},
pages = {186-194},
year = {1963},
issn = {0029-5582},
doi = {https://doi.org/10.1016/0029-5582(63)90264-5},
url = {https://www.sciencedirect.com/science/article/pii/0029558263902645},
author = {P.H. Bowen and G.C. Cox and G.B. Huxtable and J.P. Scanlon and J.J. Thresher and A. Langsford and H. Appel},
}

@article{Holt2018,
   title={Tensor Fermi liquid parameters in nuclear matter from chiral effective field theory},
   volume={97},
   ISSN={2469-9993},
   url={http://dx.doi.org/10.1103/PhysRevC.97.054325},
   DOI={10.1103/physrevc.97.054325},
   pages = {054325},
   number={5},
   journal={Phys. Rev. C},
   publisher={American Physical Society (APS)},
   author={Holt, J. W. and Kaiser, N. and Whitehead, T. R.},
   year={2018},
   month=may 
}

@article{Holt2011,
   author = {J. W. Holt and N. Kaiser and W. Weise},
   doi = {10.1140/epja/i2011-11128-x},
   pages = {128},
   issn = {1434-6001},
   issue = {10},
   journal = {EPJ A},
   month = {10},
   publisher = {Springer Science and Business Media LLC},
   title = {Nuclear energy density functional from chiral two-nucleon and three-nucleon interactions},
   volume = {47},
   year = {2011}
}

@article{JWNegele1981,
   author = {J. W. Negele and K. Yazaki},
   issue = {2},
   journal = {Phys. Rev. Lett.},
   month = {7},
   pages = {71},
   title = {Mean Free Path in a Nucleus},
   volume = {47},
   year = {1981}
}

@article{Negele1972,
  title = {Density-Matrix Expansion for an Effective Nuclear Hamiltonian},
  author = {Negele, J. W. and Vautherin, D.},
  journal = {Phys. Rev. C},
  volume = {5},
  issue = {5},
  pages = {1472--1493},
  numpages = {0},
  year = {1972},
  month = {May},
  publisher = {American Physical Society},
  doi = {10.1103/PhysRevC.5.1472},
  url = {https://link.aps.org/doi/10.1103/PhysRevC.5.1472}
}

@article{Jonson2004,
   author = {Björn Jonson},
   doi = {10.1016/j.physrep.2003.07.004},
   issn = {03701573},
   issue = {1},
   journal = {Phys. Rept.},
   month = {1},
   pages = {1-59},
   title = {Light dripline nuclei},
   volume = {389},
   year = {2004}
}

@article{Kasen2017,
   author = {D. Kasen and B. Metzger and J. Barnes and E. Quataert and E. Ramirez-Ruiz},
   doi = {10.1038/nature24453},
   issn = {14764687},
   issue = {7678},
   journal = {Nature},
   month = {11},
   pages = {80-84},
   pmid = {29094687},
   publisher = {Nature Publishing Group},
   title = {Origin of the heavy elements in binary neutron-star mergers from a gravitational-wave event},
   volume = {551},
   year = {2017}
}

@article{Koning2003,
   author = {A J Koning and J P Delaroche},
   issue = {3-4},
   journal = {Nucl. Phys. A},
   pages = {231-310},
   title = {Local and global nucleon optical models from 1 keV
to 200 MeV},
   volume = {713},
   url = {https://doi.org/10.1016/S0375-9474(02)01321-0},
   year = {2003}
}

@article{Lim2017,
   author = {Yeunhwan Lim and Jeremy W. Holt},
   doi = {10.1103/PhysRevC.95.065805},
   pages = {065805},
   issn = {24699993},
   issue = {6},
   journal = {Phys. Rev. C},
   month = {6},
   publisher = {American Physical Society},
   title = {Structure of neutron star crusts from new Skyrme effective interactions constrained by chiral effective field theory},
   volume = {95},
   year = {2017}
}

@article{Machleidt2011,
title = {Chiral effective field theory and nuclear forces},
journal = {Phys. Rept.},
volume = {503},
number = {1},
pages = {1-75},
year = {2011},
issn = {0370-1573},
doi = {https://doi.org/10.1016/j.physrep.2011.02.001},
url = {https://www.sciencedirect.com/science/article/pii/S0370157311000457},
author = {R. Machleidt and D.R. Entem}
}

@article{Jeukenne1977,
   author = {J-P Jeukenne and A Lejeune and C Mahaux},
   issue = {1},
   journal = {Phys. Rev. C},
   month = {7},
   pages = {80-96},
   title = {Optical-model potential in finite nuclei from Reid's hard core interaction},
   volume = {16},
   year = {1977}
}

@article{Odell2024,
   author = {D. Odell and P. Giuliani and K. Beyer and M. Catacora-Rios and M. Y.H. Chan and E. Bonilla and R. J. Furnstahl and K. Godbey and F. M. Nunes},
   doi = {10.1103/PhysRevC.109.044612},
   pages = {044612},
   issn = {24699993},
   issue = {4},
   journal = {Phys. Rev. C},
   month = {4},
   publisher = {American Physical Society},
   title = {ROSE: A reduced-order scattering emulator for optical models},
   volume = {109},
   year = {2024}
}

@article{Perey1962,
   author = {F Perey and B Buck},
   issue = {1},
   journal = {Nucl. Phys.},
   month = {9},
   pages = {353-380},
   publisher = {~) North-Holland Publishing Co},
   title = {A NON-LOCAL POTENTIAL MODEL FOR THE SCATTERING OF NEUTRONS BY NUCLEI},
   volume = {32},
   year = {1962}
}

@article{Perrotta2025,
   author = {S. Simone Perrotta and C. D. Pruitt and O. C. Gorton and J. E. Escher},
   doi = {10.1016/j.nuclphysa.2025.123037},
   pages = {123037},
   issn = {03759474},
   journal = {Nucl. Phys. A},
   month = {5},
   publisher = {Elsevier B.V.},
   title = {Towards next-generation optical potentials for nuclear reactions and structure calculations},
   volume = {1057},
   year = {2025}
}

@article{Pruitt2022,
   author = {C. D. Pruitt and J. E. Escher and R. Rahman},
   doi = {10.1103/PhysRevC.107.014602},
   issue = {1},
   journal = {Phys. Rev. C},
   month = {11},
   pages = {014602},
   title = {Uncertainty-quantified phenomenological optical potentials for single-nucleon scattering},
   volume = {107},
   url = {http://arxiv.org/abs/2211.07741 http://dx.doi.org/10.1103/PhysRevC.107.014602},
   year = {2022}
}

@article{Ripka1963,
   author = {G Ripka},
   journal = {Nucl. Phys.},
   month = {6},
   pages = {75-85},
   publisher = {North-Holland Publishing Co},
   title = {THE NON-LOCALITY OF THE OPTICAL POTENTIAL},
   volume = {42},
   year = {1963}
}

@article{Ross2015,
   author = {A. Ross and L. J. Titus and F. M. Nunes and M. H. Mahzoon and W. H. Dickhoff and R. J. Charity},
   doi = {10.1103/PhysRevC.92.044607},
   pages = {044607},
   issn = {1089490X},
   issue = {4},
   journal = {Phys. Rev. C},
   month = {10},
   publisher = {American Physical Society},
   title = {Effects of nonlocal potentials on (p,d) transfer reactions},
   volume = {92},
   year = {2015}
}

@article{Rotureau2018,
   author = {J. Rotureau and P. Danielewicz and G. Hagen and G. R. Jansen and F. M. Nunes},
   doi = {10.1103/PhysRevC.98.044625},
   issn = {24699993},
   issue = {4},
   journal = {Phys. Rev. C},
   month = {10},
   pages = {044625},
   publisher = {American Physical Society},
   title = {Microscopic optical potentials for calcium isotopes},
   volume = {98},
   year = {2018}
}

@article{Roussel-Chomaz2011,
doi = {10.1088/1742-6596/312/8/082004},
url = {https://dx.doi.org/10.1088/1742-6596/312/8/082004},
year = {2011},
month = {sep},
publisher = {},
volume = {312},
number = {8},
pages = {082004},
author = {Roussel-Chomaz, P},
title = {Reactions with exotic nuclei: recent results and challenges},
journal = {J. Phys.: Conf. Ser.}
}

@article{SFANTONI1981,
   title = {The imaginary part of the nucleon optical potential in nuclear matter},
journal = {Phys. Lett. B},
volume = {104},
number = {2},
pages = {89-91},
year = {1981},
issn = {0370-2693},
doi = {https://doi.org/10.1016/0370-2693(81)90565-7},
url = {https://www.sciencedirect.com/science/article/pii/0370269381905657},
author = {S. Fantoni and B.L. Friman and V.R. Pandharipande}
}

@article{Timofeyuk2013,
   author = {N. K. Timofeyuk and R. C. Johnson},
   doi = {10.1103/PhysRevLett.110.112501},
   pages = {112501},
   issn = {00319007},
   issue = {11},
   journal = {Phys. Rev. Lett.},
   month = {3},
   title = {Nonlocality in deuteron stripping reactions},
   volume = {110},
   year = {2013}
}

@article{Titus2014,
   author = {L. J. Titus and F. M. Nunes},
   doi = {10.1103/PhysRevC.89.034609},
   pages = {034609},
   issn = {1089490X},
   issue = {3},
   journal = {Phys. Rev. C},
   month = {3},
   publisher = {American Physical Society},
   title = {Testing the Perey effect},
   volume = {89},
   year = {2014}
}

@article{Titus2016,
   author = {L. J. Titus and F. M. Nunes and G. Potel},
   doi = {10.1103/PhysRevC.93.014604},
   pages = {014604},
   issn = {24699993},
   issue = {1},
   journal = {Phys. Rev. C},
   month = {1},
   publisher = {American Physical Society},
   title = {Explicit inclusion of nonlocality in (d,p) transfer reactions},
   volume = {93},
   year = {2016}
}

@article{Weppner2009,
   author = {S. P. Weppner and R. B. Penney and G. W. Diffendale and G. Vittorini},
   doi = {10.1103/PhysRevC.80.034608},
   pages = {034608},
   issn = {1089490X},
   issue = {3},
   journal = {Phys. Rev. C},
   month = {9},
   publisher = {American Physical Society},
   title = {Isospin dependent global nucleon-nucleus optical model at intermediate energies},
   volume = {80},
   year = {2009}
}

@article{Whitehead2021,
   author = {T. R. Whitehead and Y. Lim and J. W. Holt},
   doi = {10.1103/PhysRevLett.127.182502},
   pages = {182502},
   issn = {10797114},
   issue = {18},
   journal = {Phys. Rev. Lett.},
   month = {10},
   pmid = {34767381},
   publisher = {American Physical Society},
   title = {Global Microscopic Description of Nucleon-Nucleus Scattering with Quantified Uncertainties},
   volume = {127},
   year = {2021}
}

@article{Whitehead2019,
   author = {T. R. Whitehead and Y. Lim and J. W. Holt},
   doi = {10.1103/PhysRevC.100.014601},
   pages = {014601},
   issn = {24699993},
   issue = {1},
   journal = {Phys. Rev. C},
   month = {7},
   publisher = {American Physical Society},
   title = {Proton elastic scattering on calcium isotopes from chiral nuclear optical potentials},
   volume = {100},
   year = {2019}
}

@article{Whitehead2020,
   author = {T. R. Whitehead and Y. Lim and J. W. Holt},
   doi = {10.1103/PhysRevC.101.064613},
   pages = {064613},
   issn = {24699993},
   issue = {6},
   journal = {Phys. Rev. C},
   month = {6},
   publisher = {American Physical Society},
   title = {Neutron elastic scattering on calcium isotopes from chiral nuclear optical potentials},
   volume = {101},
   year = {2020}
}

@article{Idini2017,
   title={Ab Initio Optical Potentials and Nucleon Scattering on Medium Mass Nuclei},
   volume={48},
   ISSN={1509-5770},
   url={http://dx.doi.org/10.5506/APhysPolB.48.273},
   DOI={10.5506/aphyspolb.48.273},
   number={3},
   journal={Acta Phys. Pol. B},
   publisher={Jagiellonian University},
   author={Idini, A. and Barbieri, C. and Navrátil, P.},
   year={2017},
   pages={273} }

@article{Rotureau2017,
   author = {J. Rotureau and P. Danielewicz and G. Hagen and F. M. Nunes and T. Papenbrock},
   doi = {10.1103/PhysRevC.95.024315},
   issn = {24699993},
   issue = {2},
   journal = {Phys. Rev. C},
   month = {2},
   pages = {024315},
   publisher = {American Physical Society},
   title = {Optical potential from first principles},
   volume = {95},
   year = {2017}
}

@article{Svensson2022,
   author = {I. Svensson and A. E. and C. Forssén},
   doi = {10.1103/PhysRevC.105.014004},
   pages = {014004},
   issn = {24699993},
   issue = {1},
   journal = {Phys. Rev. C},
   month = {1},
   publisher = {American Physical Society},
   title = {Bayesian parameter estimation in chiral effective field theory using the Hamiltonian Monte Carlo method},
   volume = {105},
   year = {2022}
}

@article{Furnstahl2015,
   author = {R. J. Furnstahl and N. Klco and D. R. Phillips and S. Wesolowski},
   doi = {10.1103/PhysRevC.92.024005},
   issue = {2},
   journal = {Phys. Rev. C},
   month = {6},
   pages = {23},
   title = {Quantifying truncation errors in effective field theory},
   volume = {92},
   url = {http://arxiv.org/abs/1506.01343 http://dx.doi.org/10.1103/PhysRevC.92.024005},
   year = {2015}
}

@article{Young2003,
   author = {R D Young and D B Leinweber and A W Thomas},
   issue = {2},
   journal = {Prog. Part. Nucl. Phys.},
   pages = {399-417},
   title = {Convergence of Chiral Effective Field Theory},
   volume = {50},
   url = {http://www.elsevicr.com/locate/npe},
   year = {2003}
}

@article{VanGoffrier2025,
   author = {G. Van Goffrier},
   doi = {10.1103/PhysRevD.111.055033},
   pages = {055033},
   issn = {24700029},
   issue = {5},
   journal = {Phys. Rev. D},
   month = {3},
   publisher = {American Physical Society},
   title = {Improved precision calculation of the 0νββ contact term within chiral effective field theory},
   volume = {111},
   year = {2025}
}

@article{Tews2025,
   author = {I. Tews and R. Somasundaram and D. Lonardoni and H. Göttling and R. Seutin and J. C. and S. Gandolfi and K. Hebeler and A. Schwenk},
   doi = {10.1103/r314-6r62},
   pages = {033024},
   issue = {3},
   journal = {Phys. Rev. Res.},
   month = {5},
   publisher = {American Physical Society (APS)},
   title = {Neutron matter from local chiral effective field theory interactions at large cutoffs},
   volume = {7},
   year = {2025}
}

@article{Tews2018,
   author = {I. Tews and L. Huth and A. Schwenk},
   doi = {10.1103/PhysRevC.98.024001},
   pages = {024001},
   issn = {24699993},
   issue = {2},
   journal = {Phys. Rev. C},
   month = {8},
   publisher = {American Physical Society},
   title = {Large-cutoff behavior of local chiral effective field theory interactions},
   volume = {98},
   year = {2018}
}

@article{Melendez2019,
   title={Quantifying correlated truncation errors in effective field theory},
   volume={100},
   ISSN={2469-9993},
   url={http://dx.doi.org/10.1103/PhysRevC.100.044001},
   DOI={10.1103/physrevc.100.044001},
   pages = {044001},
   number={4},
   journal={Phys. Rev. C},
   publisher={American Physical Society (APS)},
   author={Melendez, J. A. and Furnstahl, R. J. and Phillips, D. R. and Pratola, M. T. and Wesolowski, S.},
   year={2019},
   month=oct }

@article{Wesolowski2022,
   author = {S. Wesolowski and I. Svensson and A. Ekström and C. Forssén and R. J. Furnstahl and J. A. Melendez and D. R. Phillips},
   doi = {10.1103/PhysRevC.104.064001},
   pages = {064001},
   issue = {6},
   journal = {Phys. Rev. C},
   month = {1},
   title = {Rigorous constraints on three-nucleon forces in chiral effective field theory from fast and accurate calculations of few-body observables},
   volume = {104},
   url = {http://arxiv.org/abs/2104.04441 http://dx.doi.org/10.1103/PhysRevC.104.064001},
   year = {2022}
}

@article{Mahzoon2014,
   author = {M. H. Mahzoon and R. J. Charity and W. H. Dickhoff and H. Dussan and S. J. Waldecker},
   doi = {10.1103/PhysRevLett.112.162503},
   issn = {10797114},
   issue = {16},
   journal = {Phys. Rev. Lett.},
   month = {4},
   pages = {162503},
   publisher = {American Physical Society},
   title = {Forging the link between nuclear reactions and nuclear structure},
   volume = {112},
   year = {2014}
}

@manual{stahulak_2025_17102397,
  author       = {Stahulak, Laina and
                  Holt, Jeremy},
  title        = {Research Data Used in Manuscript "Nonlocal nucleon-nucleus optical potentials from chiral effective field theory"},
  month        = sep,
  year         = 2025,
  publisher    = {Zenodo},
  doi          = {10.5281/zenodo.17102397},
  url          = {https://doi.org/10.5281/zenodo.17102397},
  note = {doi: 10.5281/zenodo.17102397}
}

\end{document}